\documentclass{article}

\usepackage[english]{babel}
\usepackage[letterpaper,top=2cm,bottom=2cm,left=3cm,right=3cm,marginparwidth=1.75cm]{geometry}

\usepackage{amsfonts} 
\usepackage{authblk}
\usepackage{amsmath}
\usepackage{graphicx}
\usepackage{braket}
\usepackage[colorlinks=true, allcolors=blue]{hyperref}

\usepackage{tikz}
\usepackage{float}
\usepackage{amsthm}
\usepackage{multirow}
\usepackage{tabularx}
\usepackage{booktabs}
\usepackage{caption}
\usepackage{subcaption}
\usepackage{cleveref}

\usepackage{cite} 

\newenvironment{acknowledgments}{\section*{Acknowledgments}}{}

\newcommand{\circlewith}[2][0.4cm]{%
  \tikz[baseline=-0.5ex] 
    \node[draw,circle,inner sep=0pt,minimum size=#1,scale=#1/0.4cm] {#2};}
\newcommand{\doticon}{%
  \tikz[baseline=-0.5ex] \fill (0,0) circle (1.8pt);%
}

\newcommand{\goto}{\rightarrow}
\def\ha{{\hat{a}}} 

\def\<{\langle}
\def\>{\rangle}
\newcommand{\nn}{\nonumber}

\newcommand{\red}[1]{#1}
\newcommand{\hl}[1]{#1}
\newcommand{\hlj}[1]{#1}

\title{Heralded Linear Optical Generation of  Dicke States}

\author[1]{Minhyeok Kang\textsuperscript{\S}}
\author[1]{Jaehee Kim\textsuperscript{\S}}
\affil[1]{SKKU Advanced Institute of Nanotechnology (SAINT), Sungkyunkwan University, Suwon 16419, Korea}
\author[2]{William J. Munro}
    \author[2]{Seungbeom Chin \thanks{sbthesy@gmail.com}}
 \affil[2]{Okinawa Institute of Science and Technology Graduate University, Okinawa 904-0495, Japan}
 \author[3, 4]{Joonsuk Huh \thanks{joonsukhuh@yonsei.ac.kr}}
\affil[3]{Department of Chemistry, Yonsei University, Seoul 03722, Republic of Korea}
\affil[4]{Department of Quantum Information, Yonsei University, Incheon 21983, Republic of Korea}

\begin{document}
\maketitle

\begingroup
\renewcommand\thefootnote{\textsuperscript{\S}}
\footnotetext{These authors contributed equally.}
\endgroup

\begin{abstract}
Entanglement is a fundamental feature of quantum mechanics and a key resource for quantum information processing. Among multipartite entangled states, Dicke states $|D_n^k\rangle$ are distinguished by their permutation symmetry, which provides robustness against particle loss and enables applications for quantum communication and computation.  
Although Dicke states have been realized in various platforms, most optical implementations rely on postselection, which destroys the state upon detection and prevents its further use. A heralded optical scheme is therefore highly desirable. Here, we present a linear-optical heralded scheme for generating arbitrary Dicke states $|D_n^k\rangle$ with $3n+k$ photons  through the framework of the linear quantum graph (LQG) picture. By mapping the scheme design into the graph-finding problem, and exploiting the permutation symmetry of Dicke states, we overcome the structural complexity that has hindered previous approaches. Our results provide a resource-efficient pathway toward practical heralded preparation of Dicke states for quantum technologies.
\end{abstract}

\section{Introduction}
\label{introduction}

Entanglement is a fundamental feature of quantum mechanics, representing correlations with no classical reference. It not only plays a central role in deepening our understanding of the foundations of quantum theory, but also serves as the critical resource behind the advantages of quantum information processing~\cite{horodecki2009quantum}. Multipartite entanglement extends quantum correlations across many systems, enabling richer structures and more powerful applications than bipartite entanglement~\cite{walter2016multipartite}. Such states are indispensable in areas including quantum cryptography~\cite{ekert1991quantum, proietti2021experimental, murta2020quantum}, quantum teleportation~\cite{bennett1993teleporting, luo2019quantum, hermans2022qubit}, quantum dense coding~\cite{bennett1992communication, guo2019advances, hu2018beating}, quantum error correction~\cite{al2023suppressing}, and quantum computation~\cite{bluvstein2024logical, madsen2022quantum}. Over the past two decades, multipartite entanglement has been realized in diverse physical platforms---from trapped ions~\cite{leibfried2005creation,hume2007high,monz201114} and neutral atoms~\cite{mandel2003controlled,bernien2017probing,omran2019generation} to superconducting circuits~\cite{neeley2010generation,dicarlo2010preparation,song201710}, and photonic systems\cite{Bouwmeester1999observation,Lu2007experimental,chin2021graph,lee2022entangling}---but scalable and resource-efficient generation of large entangled states still remains a central challenge.

Different classes of genuinely multipartite entangled states are characterized by distinct structures and applications~\cite{dur2000three,ma2024multipartite}.
Dicke states \(\ket{D_n^k}\) form one such class, defined as the equal-amplitude superposition of all computational basis states of \(n\)-qubits with Hamming weight \(k\)~\cite{dicke1954coherence}:
\begin{align}
    \ket{D^k_n} = \sum_{\substack{\mathbf{x} \in \{0,1\}^{\otimes n} \\ \mathrm{hw}(\mathbf{\mathbf{x}}) = k}}\begin{pmatrix}
        n \\ k
    \end{pmatrix}^{-1/2}\ket{\mathbf{x}}.
\end{align}
Here, \(\mathrm{hw}(\mathbf{x})\)  denotes the Hamming weight of the bitstring \textbf{x}. Dicke states have a fixed  number of qubit states ($n-k$ qubits of state 0 and $k$ qubits of state 1) and are symmetric under qubit permutations, making them compactly representable within the symmetric subspace and resilient against particle loss. Their symmetry underlies applications across quantum networking~\cite{prevedel2009experimental}, high-precision metrology~\cite{toth2012multipartite,ouyang2021robust,saleem2024achieving}, 
game theory~\cite{Ozdemir2007necessary}, and error correction~\cite{Ouyang14Permutation,Aydin2024familyof}. More recently, they have been recognized as efficient building blocks for variational quantum eigensolver (VQE) ansätze~\cite{scursulim2025multiclass}, enabling accurate simulations of many-body Hamiltonians with reduced resources.

Dicke states have been demonstrated experimentally in trapped ions~\cite{linington2008robust,hume2009preparation,ivanov2013creation,lamata2013deterministic}, atomic systems~\cite{stockton2004deterministic,xiao2007generation,Shao2010deterministic}, superconducting circuits~\cite{wu2017generation} and optics~\cite{kiesel2007experimental,prevedel2009experimental,wieczorek2009experimental,zhao2011efficient, Kasture2018}. In photonics, they have been generated using diverse sources such as spontaneous parametric down-conversion (SPDC)~\cite{kiesel2007experimental,prevedel2009experimental,wieczorek2009experimental}, Cross-Kerr non-linearity~\cite{zhao2011efficient}, and single photon~\cite{Kasture2018}. 
However, most previous works on Dicke state generation have relied on postselected methods---sorting out successful outcomes by detecting all photons---which irreversibly destroys the entanglement.
Such states cannot then serve as resources in further quantum protocols. 
This limitation makes heralded optical generation of Dicke states an important open challenge.

Heralded schemes address this issue by using ancillary photons and modes as success signals, identifying valid runs without disturbing the target state. In principle, this preserves entanglement as a usable quantum resource (see Ref.~\cite{forbes2025heralded} for a recent review). Yet, designing heralded schemes is significantly more difficult than postselected ones, since the required ancillas and correlations add structural complexity. To date, heralded generation of Dicke states has therefore remained elusive.

In this work, we propose a linear-optical heralded scheme that generates arbitrary Dicke states $|D_n^k\>$ within the linear quantum graph (LQG) picture~\cite{chin2021graph,chin2024shortcut,chin2024heralded}. The LQG framework is a way to map physical components of entanglement-generating circuits into graph elements, reducing the problem of scheme design to that of graph construction. This approach has successfully yielded heralded schemes for GHZ, W, GHZ–W superpositions~\cite{chin2024shortcut,chin2024heralded}, caterpillar graph states~\cite{chin2024boson}, and qudit entangled states~\cite{chin2024exponentially,chin2024creating}. Here, we extend the LQG picture to Dicke states, embedding their intrinsic permutation symmetry directly into the graph structure. This reduces design complexity and enables systematic heralded constructions of arbitrary Dicke states. The resulting  heralded Dicke state is compatible with existing photonic quantum platforms and can be employed as a genuine quantum resource.

This work is organized as follows: Section~\ref{review} reviews the LQG picture of sculpting protocol, the central tool we employ to construct our heralded scheme. 
Section~\ref{sec:Dicke digraph} introduces the Dicke graph 
in LQG picture, which corresponds to a sculpting operator that generates $|D_n^k\>$. Section~\ref{sec:linear optical network} demonstrates how the Dicke graph 
is implemented as a linear optical circuit that generates the Dicke state  $|D_n^k\>$.
Section~\ref{conclusions} provides concluding remarks and discussions.

\section{Review: LQG picture of the sculpting protocol}\label{review}

In this section, we review the concept of the LQG picture as a tool to find heralded schemes~\cite{chin2024shortcut,chin2024heralded}. The sculpting protocol is an operational framework for the heralded generation of multipartite entangled states. The main technical challenge lies in identifying suitable sculpting operators that produce the desired entangled target state. The LQG picture maps the physical components of the protocol onto elements of balanced bipartite graphs~\cite{chin2024shortcut}, or equivalently directed graphs~\cite{chin2024boson}, thereby providing a systematic tool for searching for such operators.

\paragraph*{Sculpting protocol.---} We can generate $n$-partite entangled states in sculpting protocol by applying spatially overlapped $\hlj{n+r}$ single-boson annihilation operators (called ``sculpting operators'') to a $\hlj{2n+r}$ boson initial state, where $\hlj{r}$ is the number of ancillary modes. In our setup, each boson has an $\hlj{n+r}$-dimensional spatial state and a two-dimensional internal state, hence creation and annihilation operators are expressed as $\ha_{j,s}^\dagger$ and $\ha_{j,s}$ ($j\in \{1,2,\cdots, \hlj{n+r}\}$, $s \in \{+,-\}$) respectively. Here we choose the computational basis in the $\mathrm{x}$-direction for later convenience, which is expressed in the $\mathrm{z}$-directional computation basis $\{|0\>, |1\>\}$ as $|\pm\> =  (|0\>\pm |1\>)/\sqrt{2}$.

We consider a setup with $n$ qubit systems and $\hlj{r}$ ancillary systems.
The initial state is given by
\begin{align}
\label{eqn:initial state}
 |\Psi_{\mathrm{init}}\>_{n,\hlj{r}}  =  \prod_{j=1}^{n}\hat{a}^{\dagger}_{j,+}\hat{a}^{\dagger}_{j,-}\prod_{l=1}^{\hlj{r}}\hat{a}^{\dagger}_{n+l,+}\ket{\rm{vac}},
\end{align} i.e., the $n$ parties contain two bosons while $\hlj{r}$ parties contain one boson.
We apply 
a sculpting operator \(\hat{A}_{\hlj{n+r}}\) that subtracts \(\hlj{n+r}\) bosons from $|\Psi_{\mathrm{init}}\>_{n,\hlj{r}}$. The sculpting operator \(\hat{A}_{\hlj{n+r}}\) has the following form:

\begin{align}
\label{eqn:sculpting operator}
    \hat{A}_{\hlj{n+r}} = \prod_{\hl{m}=1}^{\hlj{n+r}}\hat{A}^{(\hl{m})}\equiv \prod_{\hl{m}=1}^{\hlj{n+r}}\Big(\sum_{j=1}^{n}(\alpha^{(\hl{m})}_{j}\hat{a}_{j,0}+\beta^{(\hl{m})}_{j}\hat{a}_{j,1})+\sum_{l=1}^{\hlj{r}}\gamma^{(\hl{m})}_{l}\hat{a}_{n+l,+}\Big)
\end{align}
with
$ \sum_{j=1}^{n}(|\alpha^{(\hl{m})}_{j}|^2+|\beta^{(\hl{m})}_{j}|^2)+\sum_{\hl{l}=1}^{\hlj{r}}|\gamma^{(m)}_{l}|^2 = 1.$
For the above sculpting operator to generate a multipartite qubit state, the sculpting protocol must satisfy the \emph{no-bunching restriction}~\cite{chin2024shortcut}, \red{by which exactly one boson must be removed from each spatial mode by the operation of $\hat{A}_{\hlj{n+r}}$}\footnote{Interpreting this restriction in the heralded schemes, one boson remaining in the spatial mode encodes the qubit information, while the subtracted boson serves as the heralding signal.}. 
Once we find a sculpting operator that generates an entangled state, it can be automatically implemented in linear optics by translation rules introduced in~\cite{chin2024heralded} \red{(see Appendix~\ref{appendix:translation} for a brief explanation).}

\paragraph{LQG picture of sculpting protocols.---}
We can represent sculpting operators as graphs in the LQG picture, \red{which replaces the operator-finding problem with the graph-finding problem.} 
 The same sculpting operator in the LQG picture can be described by two representations: the \emph{undirected bigraph representation} (denoted as $G_{ub}$) and \emph{directed unipartite graph representation} (denoted as $G_{du}$)~\cite{chin2024boson}. Table~\ref{table:correspondence_comparison} displays the correspondence relations between the physical elements of the sculpting operator and the graph elements.

\begin{table}[b]
\caption{Correspondence relations between a sculpting operator in bosonic systems and graphs in LQG picture. Since we work in the two computational bases $\{|0\>,|1\>\}$ and $\{|+\>,|-\>\}$ ($|\pm\> = (|0\> \pm |1\>)/\sqrt{2}$), the qubit states are represented as colors $\{\text{Red, Blue}\}$ and $\{\text{Solid Black, Dashed Black}\}$. 
}
\label{table:correspondence_comparison}
\centering
\begin{tabularx}{\textwidth}{ |X | X | X  | }
\hline
\textbf{Bosonic Systems with Sculpting Operator} & \textbf{Bipartite Graph $G=(U\cup V,E)$} & \textbf{Directed Graph $G=(W,E)$} \\ 
\hline
\hline
Spatial modes $j$ & Labelled circles (\circlewith{$j$}) & Labelled circles (\circlewith{$j$}) \\ 
\hline
$\hlj{\hat{A}^{(m)}}$ & Unlabelled dots (\doticon) & Labelled circles (\circlewith{$\hlj{m}$}) \\
\hline
Spatial distributions of $\hat{A}_{i}$ & Undirected edges $\in E$ & Directed edges $\in E$ \\
\hline
Probability amplitude $\hlj{\alpha_{j}^{(m)}}$, $\hlj{\beta_{j}^{(m)}}$, $\hlj{\gamma_{j}^{(m)}}$ & Edge weight $\hlj{\alpha_{j}^{(m)}}$, $\hlj{\beta_{j}^{(m)}}$, $\hlj{\gamma_{j}^{(m)}}$ & Edge weight $\hlj{\alpha_{j}^{(m)}}$, $\hlj{\beta_{j}^{(m)}}$, $\hlj{\gamma_{j}^{(m)}}$\\ 
\hline
Qubit state $\psi$ & Edge weight $\psi$ (or color) & Edge weight $\psi$ (or color)
 \\ 
\hline
\end{tabularx}
\end{table}

 First, in the undirected bigraph representation $G_{ub}$, a sculpting operator is represented as an undirected balanced bipartite graph, which we call \emph{sculpting bigraph}. For a sculpting operator to satisfy the no-bunching restriction, the final state must be completely determined by the superposition of the perfect matchings\footnote{A perfect matching of a graph $G=(V,E)$ is a subset of edges $M \subseteq E$ such that every vertex in $V$ is incident to exactly one edge in $M$.} of the corresponding sculpting bigraph~\cite{chin2024shortcut}. 
 Ref.~\cite{chin2024shortcut} proposed a convenient subset of bigraphs---named \emph{effective perfect matching bigraphs (EPM bigraphs)}---which inherently satisfy the no-bunching restriction by construction~\cite{chin2024shortcut}. A bigraph is an EPM bigraph if all its edge-to-circle attachments match one of the configurations shown below:
    \begin{align}\label{epm_bi}
            \includegraphics[width=0.5\linewidth]{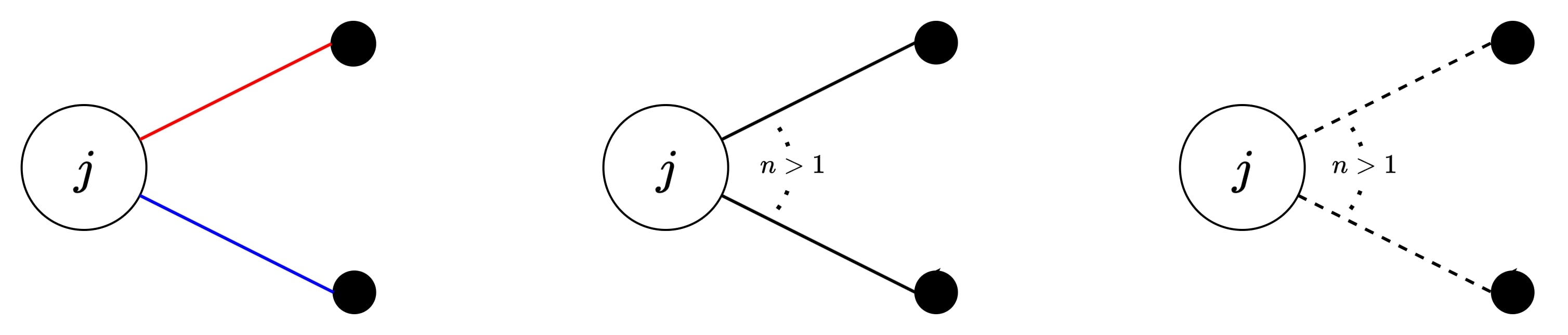},
    \end{align}
where the edge colors $\{\text{Solid Black}, \text{Dashed Black}, \text{Red} ,\text{Blue}\}$ represent internal states $\{\ket{+}, \ket{-}, \ket{0}, \ket{1}\}$ ($|\pm\> = (|0\> \pm |1\>)/\sqrt{2}$).
For an EPM bigraph in the LQG picture, we can show that \emph{the final state after the application of the corresponding sculpting operator is represented as the superposition of all the perfect matchings of the EPM bigraph}. \red{}{All graph solutions in previous works on the LQG picture~\cite{chin2024heralded,chin2024boson} correspond to EPM bigraphs, as they automatically satisfy the no-bunching constraint, a major technical challenge in identifying suitable sculpting operators.}

Second, in the directed unipartite representation $G_{du}$, the same operator is represented as a directed graph, which we call \emph{sculpting digraph}. 
A key concept in the digraph representation is the \emph{disjoint cycle cover}\footnote{A disjoint cycle cover of a digraph $G$ is a collection of directed cycles such that every vertex of $G$ lies in exactly one cycle. 
} \emph{(DCC), which corresponds to a perfect matching in $G_{ub}$}~\cite{tutteshort1954, chin2021graph, chin2024boson}. The final state is hence expressed as a superposition of all possible disjoint cycle covers in $G_{du}$. EPM bigraphs in $G_{ub}$ are mapped to a special set of digraphs, 
of which edges are connected to circles as one of the following forms: 
\begin{align}  
   \label{epm_di}
        \includegraphics[width=0.5\linewidth]{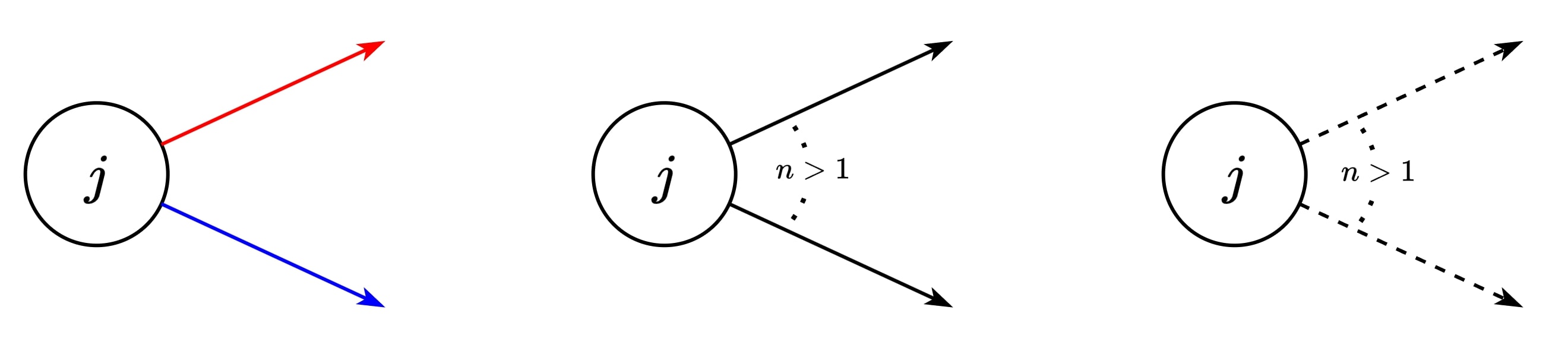}
\end{align}
While $G_{ub}$ is advantageous for intuitively illustrating sculpting protocols, $G_{du}$ is more appropriate for visualizing complex entanglement structures, as we will see in the next section. \red{The simplest Bell state generation example is given in Appendix~\ref{appendix:Bell} as proof of concept.}

\section{Dicke digraph $D_n^k$}
\label{sec:Dicke digraph}

In this section, we introduce the \emph{Dicke digraph} $D_n^k$, which corresponds to a sculpting operator for generating Dicke states. We present it in the digraph representation $G_{du}$, because it directly reveals the symmetry of the operator in its structure.

\begin{figure}[t]

    \centering

    \includegraphics[width=0.9\linewidth]{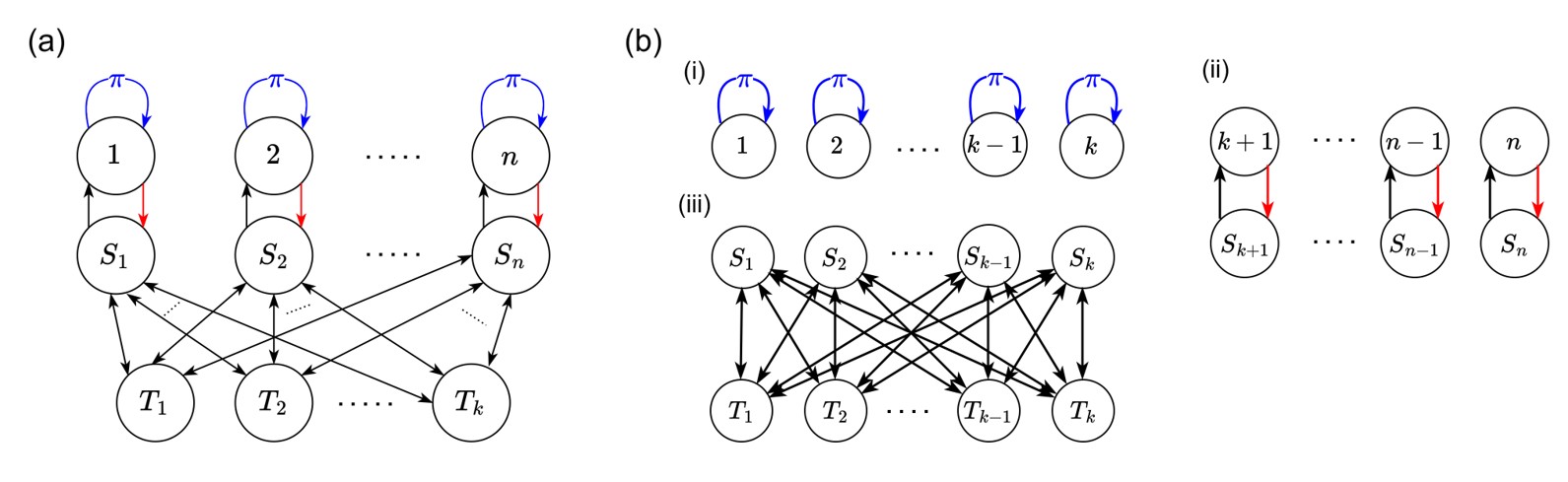}

    \caption{Dicke digraph $D^k_n$ and its Directed Cycle Covers (DCCs). 
\textbf{(a)} The Dicke digraph $D^k_n$. Edge weights are omitted, implying that all outgoing edges from a vertex have equal amplitude weights. Two-headed arrows for $S_s \leftrightarrow A_a$ represent the combination of the two directed edges $S_s \rightarrow A_a$ and $A_a \rightarrow S_s$. 
\textbf{(b)} DCCs of $D_n^k$ shown in (a). These include three types of cycles: (i) a system self-loop $(j \to j)$; (ii) a 2-cycle on a system–ancilla pair $(S_j \to j \to S_j)$; or (iii) an alternating $S\leftrightarrow T$ cycle of even length. The alternating cycles arise from the remaining vertices $U$ and $V$ forming a complete directed bipartite subgraph, where each choice of perfect matchings $S \to T$ and $T \to S$ specifies an alternating cycle cover.}

\label{fig:dicke_digraph_combined}

\end{figure}

Dicke digraph \(D_n^k\) is an EPM digraph that consists of \(n\) system vertices (denoted as $\{1,2,\cdots, n\}$)
and \(n+k\) ancillary vertices (denoted as \(S_j\) and \(T_l\), where \(j = 1,2,\dots,n\) and \(l = 1,2,\dots,k\)); \hlj{that is, $r=n+k$ in the notation of Section~\ref{review}}. We define two sets of ancillary vertices as $S\equiv \{S_j\}_{j=1}^n$ and $T\equiv \{T_l\}_{l=1}^{k}$.  The ancillary vertices form a complete directed bipartite graph between $S$ and $T$. And each system vertex \(j\) are connected to the corresponding ancillary vertex \(S_j\).
The most general form of $D_n^k$ is shown in Fig.~\ref{fig:dicke_digraph_combined} (a).
A key property of $D_n^k$ is its symmetry: permutations of the pairs  $(j,S_j )$ among different $j$ or of the  $T_l$ among $l$ leave the graph invariant. Consequently, all disjoint cycle covers of  $D_n^k$  inherit this symmetry, ensuring that the generated quantum state shares the permutation symmetry of the Dicke state $|D_n^k\>$.

\paragraph*{From Dicke graph to Dicke state.---}
From the structure of $D_n^k$, we can show that the final state generated by the corresponding sculpting operator is the Dicke state $|D_n^k\>$. 
To demonstrate this,  we first check from Fig.~\ref{fig:dicke_digraph_combined} (a) that the Dicke digraph \(D_n^k\) is an EPM digraph (see \eqref{epm_di}), hence the final state is the superposition of all its DCCs. 
Every cycle in \(D_n^k\) falls into one of three classes: a system self-loop \((j \goto j)\), a 2-cycle \(( j \goto S_j \goto j)\), or a cycle alternating through $S$ and $T$, i.e., \(S\leftrightarrow T\). Since system vertices are only connected to \(\{S_j\}_{j=1}^n\) or itself, any cycle through \(j\) is either \((j \goto j)\) or \(( j \goto S_j \goto j)\); and each \(T_l\) is only connected to \(S\), hence any cycle including \(T\) must be in \(S\leftrightarrow T\). 
Then we can see that \emph{all the DCCs contain $n-k$ self-loops} $(j \goto j)$, which is straightforward given that the elements in \(T\) are only connected to those in \(S\) and hence $k$ elements in  \(S\) are always included in the cycle $S\leftrightarrow T$.   



Now let us consider the case when the system vertices $1,2, \cdots, k$ of the DCC have self-loops (see Fig.~\ref{fig:dicke_digraph_combined} (b)). Then the other system vertices $k+1, k+2,\cdots, n$ are automatically included in \(( j \goto S_j \goto j)\). 
The remaining cycles are those alternating between \(U=\{S_1,\cdots, S_k\}\) and \(V=\{T_1,\dots,T_k\}\). Since the subgraph including $S$ and $T$ constructs a complete balanced bipartite directed graph, we can see that it has $(k!)^2$ distinct DCCs in it. 
From Table~\ref{table:correspondence_comparison}, we can directly see that these $(k!)^2$ DCCs with the same self-loops correspond to the same operator. 

Indeed, the DCCs of the subgraph in Fig.~\ref{fig:dicke_digraph_combined} (b) are all written in the operator form as
\begin{align}
\Big(\prod_{m=1}^{k}-\hat{a}_{m,1}\Big)
\Big(\prod_{p=k+1}^{n}\hat{a}_{p,0}\Big)
\Big(\prod_{q =1}^{n}\hat{a}_{S_q, +}\Big)
\Big(\prod_{l=1}^{k}\hat{a}_{T_l, +}\Big).
\end{align}
This operator is applied to the $(3n+k)$-particle initial state $|\Psi_{\mathrm{init}}\>_{n,n+k}$ to obtain the state
\begin{align}
 \Big(\prod_{m=1}^{k}\hat{a}^\dagger_{m,1}\Big)\Big(\prod_{p=k+1}^{n}\hat{a}^\dagger_{p,0}\Big)|\text{vac}\> = |\underbrace{1,1,\cdots, 1}_{k},\underbrace{0,0,\cdots, 0}_{n-k}\>,
\end{align}
which is directly derived using the identity
\begin{align}\label{qubit_identities}
		\ha_{j,0}\ha^\dagger_{j,+}\ha^\dagger_{j,-}|\mathrm{vac}\> = \ha^\dagger_{j,0}|\mathrm{vac}\>, \quad \ha_{j,1}\ha^\dagger_{j,+}\ha^\dagger_{j,-}|\mathrm{vac}\> = - \ha^\dagger_{j,1}|\mathrm{vac}\>. 
\end{align} 
By the permutation symmetry among the exchange of system qubits, the total final state becomes proportional to $|D_n^k\>$. By recovering all the normalization factors, the final state $|\Psi_{\rm{fin}}\>$ is given by
\begin{align}
    \label{eqn:EPM dicke state}
  |\Psi_{\rm{fin}}\> = \hat{A}_{2n+k}|\Psi_{\rm{init}}\>  \sim 
  \ket{D^k_n}\ket{\rm{vac}}.
\end{align}

\begin{figure}[t]
    \centering
    \includegraphics[width=1\linewidth]{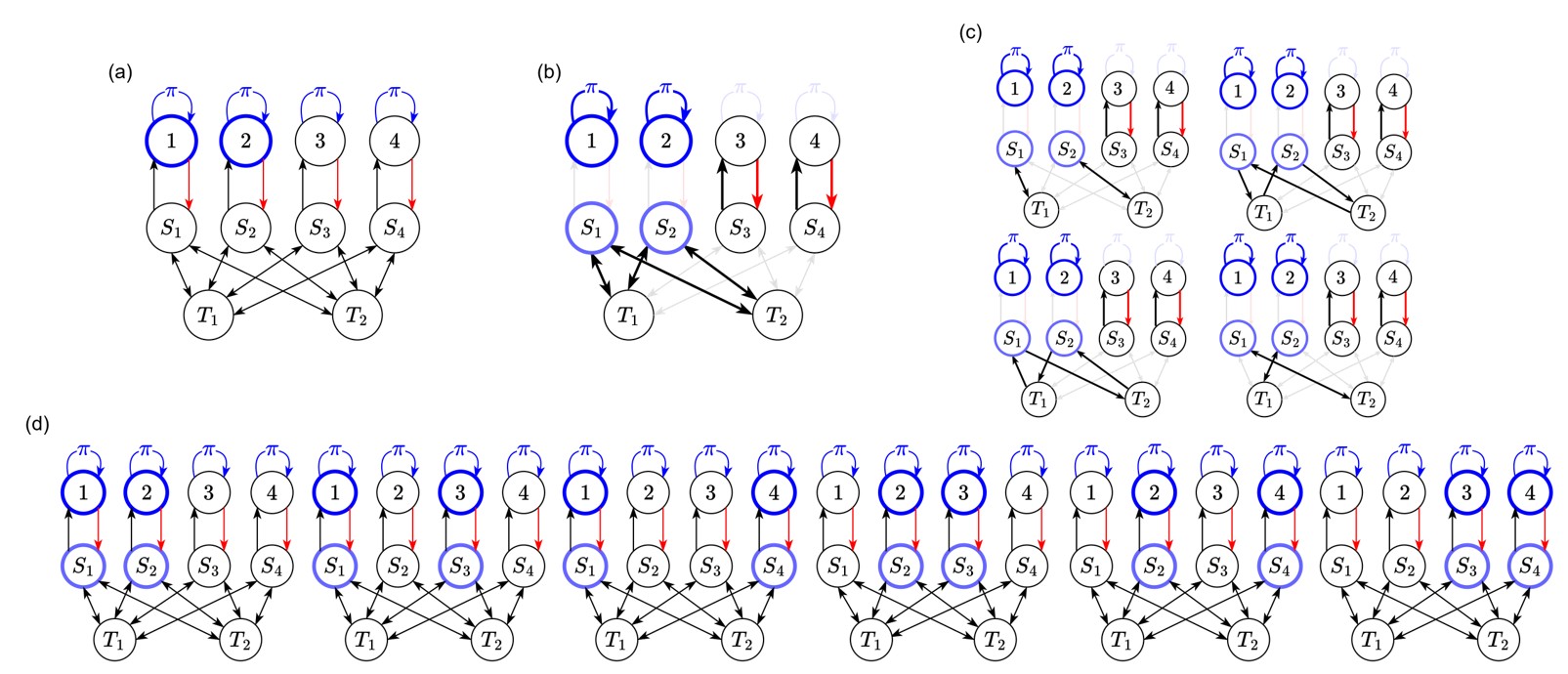}
    \caption{Example of the mechanism for the Dicke digraph with \((n,k)=(4,2)\). 
(a) 
(b) The corresponding system vertices take self-loops, while the remaining system vertices form 2-cycles. 
\(U=\{S_1,S_2\}\) and \(V=\{T_1,T_2\}\) form a complete balanced directed bipartite subgraph. 
(c) Four disjoint cycle covers are possible in the subgraph, which correspond to the same operator monomial and generate $|1100\>$. 
(d) By the permutation symmetry among $(j,S_j)$ ($j\in\{1,2,3,4\}$) of the Dicke digraph, we can see that there are six more subgraphs that give permuted states of $|1100\>$, hence Dicke state $|D_4^2\>$.
}
    \label{fig:temporary}
\end{figure}

\paragraph{$D_4^2$ example.---} 
As a proof of concept, we analyze the simplest non-trivial $(n,k) = (4,2)$ case here (see Fig.~\ref{fig:temporary}). 
First, we consider DCCs that have self-loops on 1 and 2 (Fig.~\ref{fig:temporary} (a)). 
Then the other system vertices 3 and 4 pair with their corresponding ancillas to form the 2-cycles \((S_3 \goto 3 \goto S_3)\) and \((S_4 \goto 4 \goto S_4)\) (Fig.~\ref{fig:temporary} (b)). 
Now the remaining vertices are \(U=\{S_1,S_2\}\) and \(V=\{T_1,T_2\}\); these induce a complete balanced bipartite directed graph. 
In \(U \cup V\), a disjoint cycle cover is obtained by choosing one perfect matching for \(S \to T\) and another independent perfect matching for \(T \to S\), giving \((2!)^2=4\) possibilities (Fig.~\ref{fig:temporary} (c)). 
These four DCCs all yield the same system operator monomial
\begin{align}
\Big(\prod_{m=1}^{2}-\hat{a}_{m,1}\Big)
\Big(\prod_{p=3}^{4}\hat{a}_{p,0}\Big)
\Big(\prod_{q =1}^{4}\hat{a}_{S_q, +}\Big)
\Big(\prod_{l=1}^{2}\hat{a}_{T_l, +}\Big).   
\end{align}

Applying the above operator to the initial state 
\begin{align}
    |\Psi_{\rm{init}}\>_{4,6} =  \prod_{j=1}^{4}\hat{a}^{\dagger}_{j,+}\hat{a}^{\dagger}_{j,-}\prod_{l=1}^{4}\hat{a}^{\dagger}_{S_{l, +}}\prod_{m=1}^{2}\hat{a}^{\dagger}_{T_{m, +}}  \ket{\rm{vac}},
\end{align}
we obtain \(|1100\rangle\). 
By the permutation symmetry among $(j,S_j)$ ($j\in\{1,2,3,4\}$) of the Dicke digraph, we can see that the final state is a superposition of the qubit permutation of $|1100\>$, hence becomes the Dicke state $|D_4^2\>$ (Fig.~\ref{fig:temporary} (d)). \red{With this graph solution, we generate $|D_4^2\>$ by subtracting 10 particles from the initial 14 particles.}

\begin{figure}[t]
    \centering
    \includegraphics[width=0.9\textwidth]{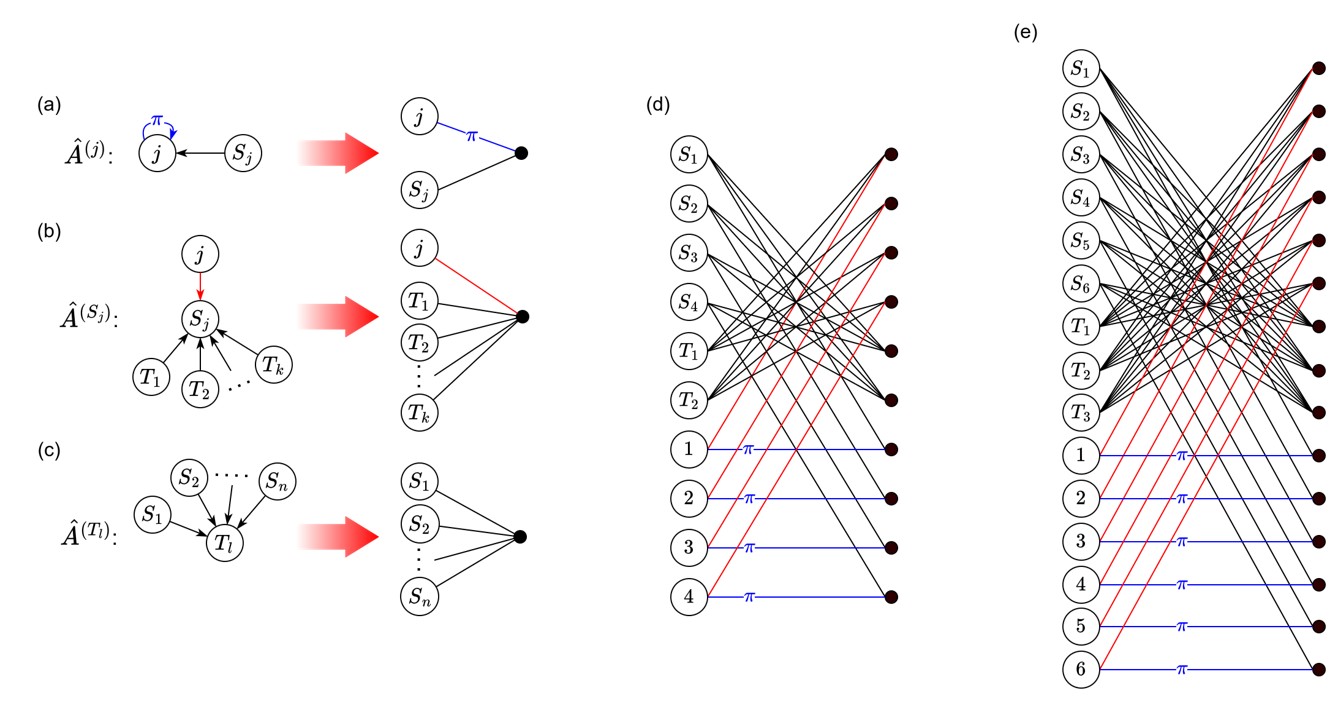}
    \caption{Conversion of a Dicke digraph into a sculpting bigraph. (a--c) Local subgraph transformation from $G_{du}$ to $G_{ub}$. 
    Node indices, colors and edge weights (including signs) are preserved. (d) Bigraphs \(D^2_4\) and (e) \(D^3_6\) constructed from the conversion. } 
    \label{fig:conversion}
\end{figure}

\begin{figure}[t]
    \centering
    \includegraphics[width=0.8\linewidth]{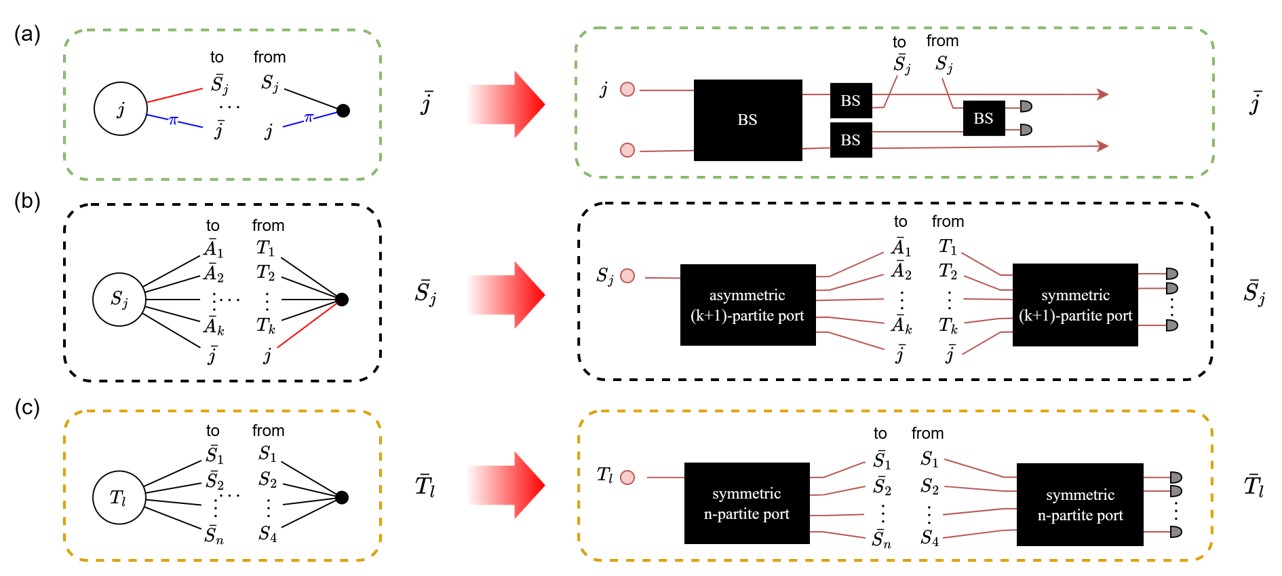}
    \caption{Adopted translation rules from bigraphs to linear optical networks based on Fig.~\ref{fig:translation_rules_dualrail}. We decomposed $D_n^k$ into circle-dot pairs on the same line in the bigraph. Each dashed box on the LHS is an open subgraph, whose open edges are attached to dots or circles following the designated labels. Then we obtain the circuit elements in the dotted boxes on the RHS from the translation rules in Fig. 2 of Ref.~\cite{chin2024heralded}. By connecting the open wires in the circuit elements, we can uniquely construct the linear optical circuit that generates $|D_n^k\>$ by heralding.
    }
    \label{fig:linear optical network blocks}
\end{figure}

\begin{figure}[t]
    \centering
    \includegraphics[width=.8\linewidth]{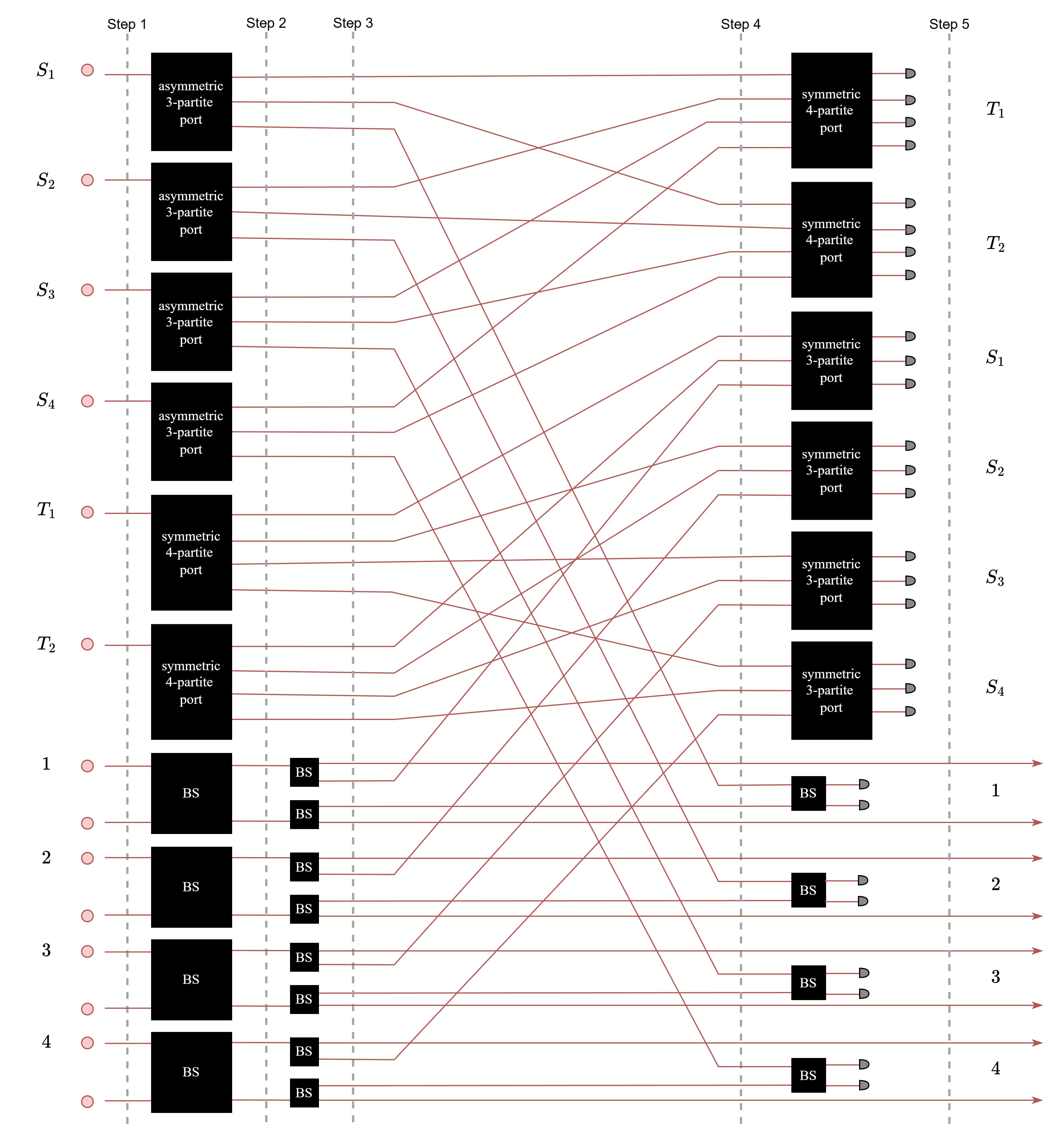}
    \caption{Linear optical network from Dicke digraph \(D_4^2\), split by several steps. Note that this figure is drawn with the positions of the $4$-partite port and the $3$-partite port swapped for clarity.}
    \label{fig:D_4_2 success prob}
\end{figure}

\section{Heralded generation of Dicke states from Dicke graphs}
\label{sec:linear optical network}

In this section, based on the structure of $D_n^k$, we provide a physical setup that generates heralded Dicke states $|D_n^k\>$. While the sculpting protocol we propose can be implemented in any bosonic system with linear operations in principle, we present a linear optical circuit in this work. 
Following the procedure in Ref.~\cite{chin2024heralded}, we design the circuit by first converting the Dicke digraph in $G_{du}$ into an EPM bigraph in $G_{ub}$, and then mapping bigraph elements to optical elements via a set of translation rules. Those elements are assembled based on the structure of $D_n^k$, yielding the heralded circuit for our target state.

The transformation of $D_n^k$ from $G_{du}$ to $G_{ub}$ is given in Fig.~\ref{fig:conversion}. The digraph $D_n^k$ is decomposed into three elements denoted as \(\hat{A}^{(j)}\), \(\hat{A}^{(S_j)}\), and \(\hat{A}^{(T_l)}\) in Fig.~\ref{fig:conversion} (a)-(c). From the second column of Table~\ref{table:correspondence_comparison}, each elements are replaced with a dot connected to the relevant circles in $G_{ub}$. Those three elements completely compose the bigraph form of $D_n^k$ for arbitrary $n$ and $k$. Fig.~\ref{fig:conversion} (d) displays the $D_4^2$ example.

Then we can construct a linear optical setup following the translation rules from the bigraph elements to linear optical elements~\cite{chin2024heralded}, \red{which is summarized in Appendix~\ref{appendix:translation}}. 
We adopt a dual-rail implementation using asymmetric and symmetric $n$-partite multiport interferometers, where the symmetric ones install the $n$-level discrete Fourier transformations (note that $n=2$ multiport interferometer is the BS). 
In this architecture, each system mode $m$ is represented as a dual-rail qubit composed of two spatial paths 
$(m,0)$ and $(m,1)$, 
corresponding to the logical states $\ket{0_L}$ and $\ket{1_L}$, respectively, while all ancillary and interface modes are treated as single-rail optical modes\footnote{Although the main derivation in the manuscript adopts this dual-rail implementation, the same construction can be equivalently realized using polarization encoding.}. 
We then translate each circle–dot pair into linear optical components as shown in Fig.~\ref{fig:linear optical network blocks}. Then the dashed boxes $\bar{j}$, $\bar{S}_j$ and $\bar{T}_l$ are connected to each other following the bigraph structure of $D_n^k$, hence we can uniquely construct a linear optical circuit for $|D_n^k\>$.
For the simplest nontrivial $|D_4^2\>$ case, the circuit is as in Fig.~\ref{fig:D_4_2 success prob}.

Now we directly check that our circuit actually generates the target state, and derive the success probability by heralding.  As denoted in Fig.~\ref{fig:D_4_2 success prob}, we divide the process into 5 steps:

$ $\\
\noindent
\textbf{Step 1.} Preparation of the initial state as  
    \begin{align} \label{eq:initial}
    \left(\prod_{j=1}^{n}\hat{a}_{S_j}^{\dagger}\right)
     \left(\prod_{l=1}^{k}\hat{a}_{T_l}^{\dagger}\right)
     \left(\prod_{m=1}^{n}\hat{a}_{m\, 0}^{\dagger}\hat{a}_{m \, 1}^{\dagger}\right).
    \end{align} 
\noindent 
\textbf{Step 2.} Division of the photon paths with BSs, asymmetric and symmetric multi-partite ports. Note that the number of paths divided from the ancillary modes corresponds to the number of edges for each ancillary vertex in the bigraph representation. The initial state as in Eq.~\eqref{eq:initial} is transformed to
    \begin{align}
    \frac{1}{2^nn^{k/2}}
    \left(\prod_{j=1}^{n}\left(\alpha \sum_{s=1}^{k}\hat{a}_{S_j\, s}^{\dagger} + \beta \hat{a}_{S_j\, k+1}^{\dagger}\right)\right)\cdot\left(\prod_{l=1}^{k}\left(\sum_{t=1}^{n}\hat{a}_{T_l \, t}^{\dagger}\right)\right) \cdot \left(\prod_{m=1}^{n}(\hat{a}_{m \, 0}^{\dagger 2}-\hat{a}_{m \, 1}^{\dagger 2})\right).
    \end{align} 
    Note that the probability amplitudes $\alpha$ and $\beta$ ($k|\alpha|^2 + |\beta|^2=1$) from asymmetric multiports are set to satisfy the symmetry of the circuit. We can control them so that the entire circuit can have the maximal success probability.
 We denote the creation operator of a mode split by the $N$-partite port or the BS from a given mode by adding additional indices, i.e., from $\hat{a}_{S_j}^{\dagger}$ to $\hat{a}_{S_j\,s}^{\dagger}$ with $s = 1,\dots,k+1$, and from $\hat{a}_{T_l}^{\dagger}$ to $\hat{a}_{T_l\,t}^{\dagger}$ with $t = 1,\dots,n$.

\noindent 
\textbf{Step 3.} Splitting two-photon states with BSs.  By denoting the operator \(\hat{a}_{j\,\sigma}^{\dagger}\) after the
BS operation
as \(\hat{a}_{j\;\sigma\;\sigma'}^{\dagger}\) with \(\sigma' = 0,1\), the state evolves into
    \begin{align}\label{eq:step3}
     \frac{1}{2^{2n}n^{k/2}} &\left( \prod_{j=1}^{n}\left( \alpha \sum_{s=1}^{k}\hat{a}_{S_j\, s}^{\dagger} + \beta \hat{a}_{S_j\, k+1}^{\dagger} \right)\right)
     \cdot \left(\prod_{l=1}^{k}\left(\sum_{t=1}^{n}\hat{a}_{T_l \,t}^{\dagger}\right)\right)\\ \nn
     &\times \left(\prod_{m=1}^{n}\left((\hat{a}_{m \, 0\,0}^{  \dagger} + \hat{a}_{m \, 0\,1}^{  \dagger})^2-(\hat{a}_{m \, 1\,0}^{  \dagger} - \hat{a}_{m \, 1\,1}^{\dagger})^2\right)\right).
    \end{align}

\noindent
\textbf{Step 4.} Permutation of wires. State~\eqref{eq:step3} evolves into 
    \begin{align}\label{eq:step4}
        \frac{1}{2^{2n}n^{k/2}} &\left(\prod_{j=1}^{n}\left(\alpha \hat{a}_{T_1\,j}^{\dagger} + \dots + \alpha \hat{a}_{T_k\,j}^{\dagger} + \beta \hat{a}_{j\,0\,1}^{\dagger}\right)\right)
        \cdot \left(\prod_{l=1}^{k}\left(\hat{a}_{S_1\,l}^{  \dagger}+ \dots + \hat{a}_{S_n\,l}^{  \dagger}\right)\right)\\ \nn 
        &\times  \left(\prod_{m=1}^{n}\left((\hat{a}_{m \, 0\,0}^{  \dagger} + \hat{a}_{S_m \, k+1}^{  \dagger})^2-(\hat{a}_{m \, 1\,0}^{  \dagger} - \hat{a}_{m \, 1\,1}^{  \dagger})^2\right)\right)
        .
    \end{align}

\noindent
\textbf{Step 5.} Application of the $n$-partite port and BS before detection. Then \eqref{eq:step4} evolves into 
\begin{align}\label{eq:final}
\frac{1}{2^{2n} n^{k/2}}
&\left(\prod_{j=1}^{n}
\left(
\alpha \sum_{p=1}^{n} (U_{n})_{p j}\,\hat{a}_{T_1\, p}^{  \dagger}
+ \cdots +
\alpha \sum_{p=1}^{n} (U_{n})_{p j}\,\hat{a}_{T_k\, p}^{  \dagger}
+ \frac{\beta}{\sqrt{2}}\left(  \hat{a}_{j\,0\,1}^{  \dagger}  + \hat{a}_{j\,1\,0}^{  \dagger}\right)
\right)
\right)
\\
&\times
\left(\prod_{l=1}^{k}
\left(
\sum_{q=1}^{k+1} (U_{k+1})_{q l}\,\hat{a}_{S_1\, q}^{  \dagger}
+ \cdots +
\sum_{q=1}^{k+1} (U_{k+1})_{q l}\,\hat{a}_{S_n\, q}^{  \dagger}
\right)
\right)
\\
&\times 
\left(
\prod_{m=1}^{n}
\left(
\left(\hat{a}_{m\,0\,0}^{ \dagger} + \sum_{q=1}^{k+1} (U_{k+1})_{q\,k+1}\,\hat{a}_{S_m\, q}^{  \dagger}\right)^2
-
\left(
\frac{1}{\sqrt{2}}\left(\hat{a}_{m\,0\,1}^{  \dagger} - \hat{a}_{m\,1\,0}^{  \dagger}\right) - \hat{a}_{m\,1\,1}^{ \dagger}
\right)^2
\right)
\right).
\end{align}
Here, $U_d$ denotes the discrete Fourier transform matrix corresponding to a symmetric $d$-partite multiport.


After the postselection, all creation operators associated with the detected (upper-most) modes are removed, and only the non-detected modes remain. 
 Collecting all $\binom{n}{k}$ possible combinations where $k$ of the $n$ modes correspond to the latter case gives the normalized post-selected state: 
\begin{equation}
\binom{n}{k}^{-1/2}
\sum_{\substack{M\subset\{1,\dots,n\}\\|M|=k}}
\left(\prod_{m\in M}\hat a^{\dagger}_{m11}\right)
\left(\prod_{m\notin M}\hat a^{\dagger}_{m00}\right)
\ket{\mathrm{vac}}
\end{equation}
with amplitude
\begin{equation}
\binom{n}{k}^{1/2}\,
\frac{k!^2}{2^{\frac{3n}{2}}\,n^{k}\,(k+1)^{\frac{n}{2}}}\,
\beta^{n-k}\alpha^k.
\end{equation}
The factor $\beta^{n-k}\alpha^k$ attains its maximum value when
\begin{equation}
\beta = \sqrt{\frac{n-k}{n}},\qquad
\alpha = \sqrt{\frac{1}{n}}.
\end{equation}

Including the feed-forward factor $2^{\,n} n (k+1)$, 
the total success probability becomes
\begin{align}
P_{\mathrm{suc}}
= \binom{n}{k}\,
\frac{(k!)^{4}(n-k)^{\,n-k}}{2^{2n}\,n^{\,n+2k-1}(k+1)^{n-1}}.
\end{align} 


\red{For $(n,k)=(4,2)$ and (5,2), the success probabilities are approximately $3.4\times 10^{-6}$ and $~1.3\times 10^{-7}$, respectively. 
These values exceed typical dark-count probabilities per detection window (of order $10^{-9}$) for modern superconducting nanowire single-photon detectors operating with Hz-level dark count rates and nanosecond timing resolution.  While the success probability decreases rapidly with increasing system size, these results indicate that proof-of-principle demonstrations of the scheme are feasible for moderate system sizes with current photonic technology. The generation rate may be further enhanced through temporal or spatial multiplexing of photon sources, parallelized detection, and fast feed-forward switching~\cite{meyer2020single,bustard2024toward}, which have been widely explored in photonic quantum information processing.  These techniques make the scheme experimentally feasible within photonic quantum technologies.
}

\red{
Finally, we comment on the effect of optical loss in the proposed scheme. In realistic photonic implementations, losses may occur in both system and ancillary modes due to imperfect optical components and detector inefficiencies. In a heralded protocol such as ours, photon loss primarily reduces the overall success probability and heralding rate, since loss events lead to failure of the required detection pattern. Crucially, because successful generation is conditioned on specific heralding outcomes, optical loss does not introduce false-positive events but instead decreases the rate of valid state preparation. A detailed quantitative analysis of loss effects, including mode-dependent transmission and detector inefficiencies, is left for future work.
}

\section{Conclusions}\label{conclusions}



\red{In this work, we have introduced a linear-optical heralded scheme for generating arbitrary Dicke states $|D^k_n\rangle$ within the linear quantum graph (LQG) framework. By embedding the permutation symmetry of Dicke states directly into the graph structure, we reduce the complexity of constructing suitable sculpting operators and provide a systematic pathway toward scalable implementations. The resulting scheme can be translated into linear optical circuits using standard components, making it compatible with existing photonic quantum technologies. To our knowledge, no prior heralded linear optical scheme for generating Dicke states  has been reported.}

\red{
We briefly compare our heralded approach with post-selected schemes for generating Dicke states. Post-selected methods typically require fewer photons and can achieve higher success probabilities, as they do not impose heralding constraints. However, such schemes are inherently destructive: the measurements used to identify successful events consume the generated state, preventing its use in subsequent quantum information processing tasks. In contrast, heralded schemes trade reduced success probability for the ability to prepare the target entangled state as a usable resource. This distinction is particularly relevant for applications in quantum networking and distributed quantum protocols, where the availability of the generated state is essential.
}

\red{We also comment on the effect of optical loss in realistic implementations. Losses may arise from imperfect optical components and detector inefficiencies in both system and ancillary modes. In a heralded protocol such as ours, photon loss primarily reduces the overall success probability and heralding rate, since loss events lead to failure of the required detection pattern. Importantly, because successful state generation is conditioned on specific heralding outcomes, loss does not introduce false-positive events but instead decreases the rate of valid state preparation. A detailed quantitative analysis of loss effects, including mode-dependent transmission and detector efficiencies, is left for future work.}

\red{Finally, we comment on the role of nonlinear interactions. Strong light–matter interfaces, such as those realized in cavity or waveguide QED systems, could in principle enable more deterministic generation of multipartite entangled states and thereby enhance generation rates. However, such approaches rely on fundamentally different physical platforms and typically introduce additional sources of noise and experimental complexity. In contrast, the present scheme operates entirely within linear optics, where probabilistic generation combined with heralding provides a practical and scalable pathway using current photonic technologies.}

\red{
Overall, our results highlight the usefulness of the LQG framework for the heralded generation of Dicke states using linear optics. We anticipate that this approach can be extended to other classes of multipartite entangled states and may serve as a useful tool for the design of photonic quantum networks and distributed quantum information protocols.}


\begin{acknowledgments}

JH acknowledges support by the National Research Foundation of Korea (NRF, RS-2023-NR068116, RS-2025-03532992), the Institute for Information \& Communications Technology Promotion (IITP) grant funded by the Korea government (MSIP) (No. 2019-0-00003), and the Yonsei University Research Fund under project number 2025-22-0140.
SC is supported by National Research Foundation of Korea (NRF, RS-2023-00245747).
WJM was partially supported by the JSPS Kakenhi Grant No. 21H04880.

\end{acknowledgments}

\appendix

\section{LQG representation for a Bell state}
\label{appendix:Bell}

\hl{In this section, we give a simple example of the LQG representation as proof of concept. Our target state is the Bell state $(\ket{01}+\ket{10})/\sqrt{2}$, which gives the simplest example of the LQG representation that can be understood as the Dicke state $\ket{D_{2}^{1}}$. We first explain the graph solution in Ref.~\cite{chin2024shortcut} without any ancillary mode and compare with the Dicke digraph $D_{2}^{1}$ that can generate the same target state.}

\begin{figure}[t]
    \centering
    \includegraphics[width=1.0\linewidth]{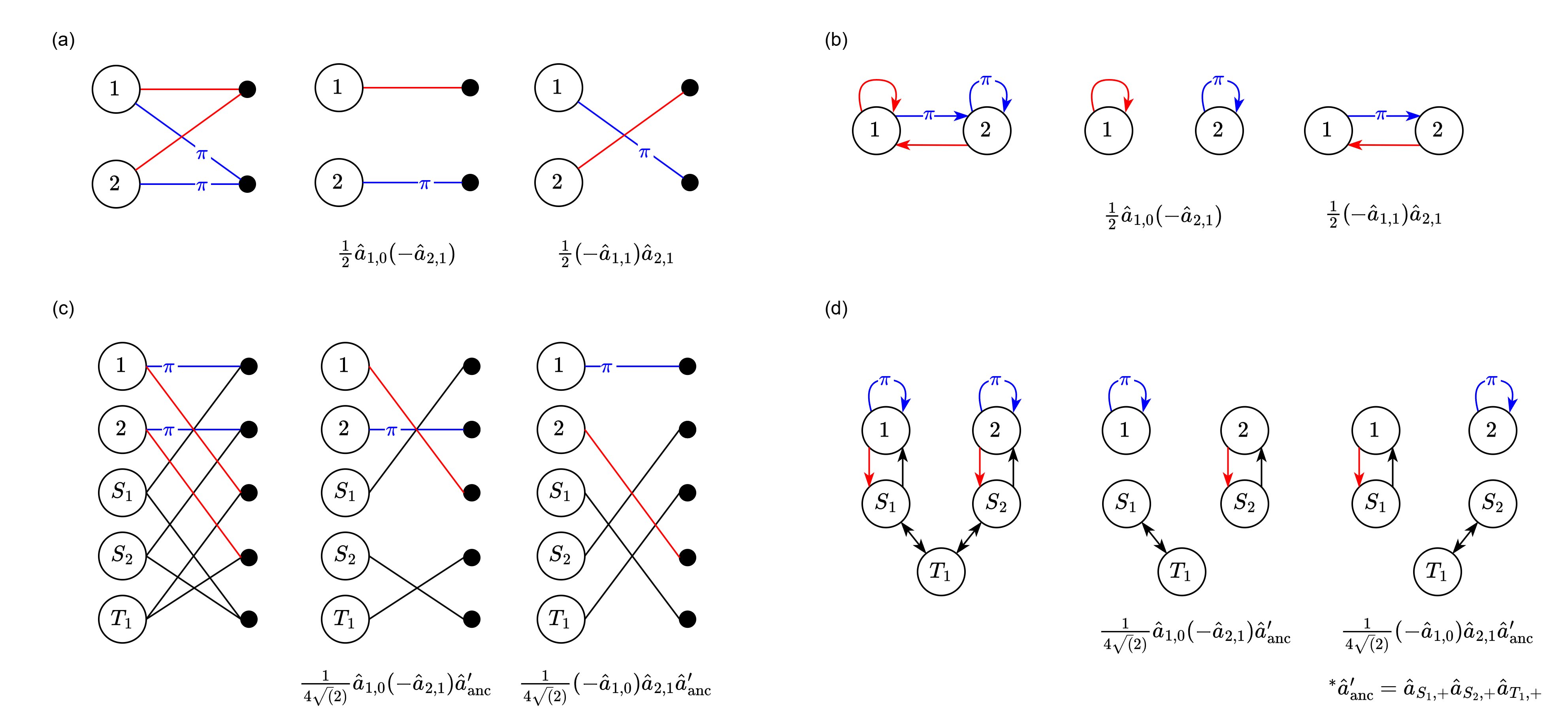}
    \caption{\hl{Sculpting bigraph and sculpting digraph for the Bell state. 
    (a) The sculpting bigraph and its perfect matchings. 
    (b) The sculpting digraph and its DCCs.
    Sculpting bigraph and sculpting digraph for the Dicke state $\ket{D_2^1}$.
    (c) The sculpting bigraph and its perfect matchings. 
    (d) The sculpting digraph and its DCCs.
    }}
    \label{fig:Bell1}
\end{figure}

\red{
First, the sculpting scheme for generating the Bell state in Ref.~\cite{chin2024shortcut} consists of two qubit systems, hence the initial state is given by
\begin{align}
    \ket{\Psi_{\rm{init}}} = \prod_{j=1}^{2}\hat{a}_{j,+}^{\dagger}\hat{a}_{j,-}^{\dagger}\ket{\rm{vac}}.
\end{align}
The sculpting operator is given by
\begin{align}
    \hat{A}_2 = \hat{A}^{(2)}\hat{A}^{(1)}, \quad
    \hat{A}^{(1)} = \frac{1}{\sqrt{2}}\left(\hat{a}_{1,0} + \hat{a}_{2,1}\right), \quad
    \hat{A}^{(2)} = \frac{1}{\sqrt{2}}\left(-\hat{a}_{1,1} - \hat{a}_{2,0}\right).
\end{align}
Applying the sculpting operator to the initial state gives
\begin{align}\label{bell_final}
    \hat{A}_2\ket{\Psi_{\rm{init}}} = \frac{1}{2}\left(\hat{a}_{1,0}^{\dagger}\hat{a}_{2,1}^{\dagger} + \hat{a}_{1,1}^{\dagger}\hat{a}_{2,0}^{\dagger}\right)\ket{\rm{vac}},
\end{align}
the Bell state. From the correspondence relations in Table~\ref{table:correspondence_comparison},  the sculpting bigraph and sculpting digraph for this protocol are illustrated in Fig.~\ref{fig:Bell1} (a) and (b), respectively. One can directly check that they satisfy EPM graph restrictions Eq.~\eqref{epm_bi} and Eq.~\eqref{epm_di}. All perfect matchings of the bigraph and all DCCs of the digraph, along with their corresponding operators, are also shown therein. 
}

\hl{Second, since the Bell state can be thought of as the Dicke state $\ket{D_{2}^1}$, it can be generated from the Dicke digraph $D_{2}^{1}$ and its corresponding sculpting bigraph as illustrated in Fig.~\ref{fig:Bell1} (c) and (d). In this case, there are three ancillary systems $S_1$, $S_2$, and $T_1$, so the initial state is
\begin{align}
    \ket{\Psi_{\rm{init}}} = \hat{a}_{T_1,+}^{\dagger}\prod_{j=1}^{2}\hat{a}_{S_j,+}^{\dagger}\hat{a}_{j,+}^{\dagger}\hat{a}_{j,-}^{\dagger}\ket{\rm{vac}}.
\end{align}
Both the bigraph and digraph satisfy the EPM conditions. All perfect matchings of the bigraph and all DCCs of the digraph, along with their corresponding operators, are also shown therein. Here we denote the product of annihilation operators for all ancillary systems as $\hat{a}'_{\rm{anc}} = \hat{a}_{S_1,+}\hat{a}_{S_2,+}\hat{a}_{T_1,+}$. The sum of operators acting on the initial state gives
\begin{align}
    \frac{1}{4\sqrt{2}}\left(-\hat{a}_{1,1}\hat{a}_{2,0}\hat{a}'_{\rm{anc}} - \hat{a}_{1,0}\hat{a}_{2,1}\hat{a}'_{\rm{anc}}\right)\ket{\Psi_{\rm{init}}} = \frac{1}{4\sqrt{2}}\left(\hat{a}_{1,1}^{\dagger}\hat{a}_{2,0}^{\dagger}+\hat{a}_{1,0}^{\dagger}\hat{a}_{2,1}^{\dagger}\right)\ket{\rm{vac}},
\end{align}
the same final state with Eq.~\eqref{bell_final}.
}



\section{Translation rules from EPM bigraphs to linear optical circuits}\label{appendix:translation}

\red{Ref.~\cite{chin2024heralded} establishes translation rules that map each elementary block of an EPM sculpting bigraph to a linear optical element with polarization encoding, which provide a direct correspondence between the bigraph description and its experimental implementation. Here we display the rules adapted to the dual-rail encoding used in this work.}

\red{
\paragraph*{Optical elements.---}
In dual-rail encoding, the translated circuits consist of the following elements:
\begin{itemize}
\item \textbf{Beam splitter (BS)}: A balanced ($50{:}50$) beam splitter that coherently mixes two input modes.
\item \textbf{$d$-partite multiport interferometer}: A linear optical network that coherently mixes $d$ input modes. In general (\emph{asymmetric}) the couplings are mode-dependent; the \emph{symmetric} multiport is the special case with uniform couplings, implementing the $d$-level discrete Fourier transform (DFT). For $d=2$, the symmetric multiport reduces to a balanced BS.
\end{itemize}
}

\begin{figure*}[t]
    \centering
    \includegraphics[width=14cm]{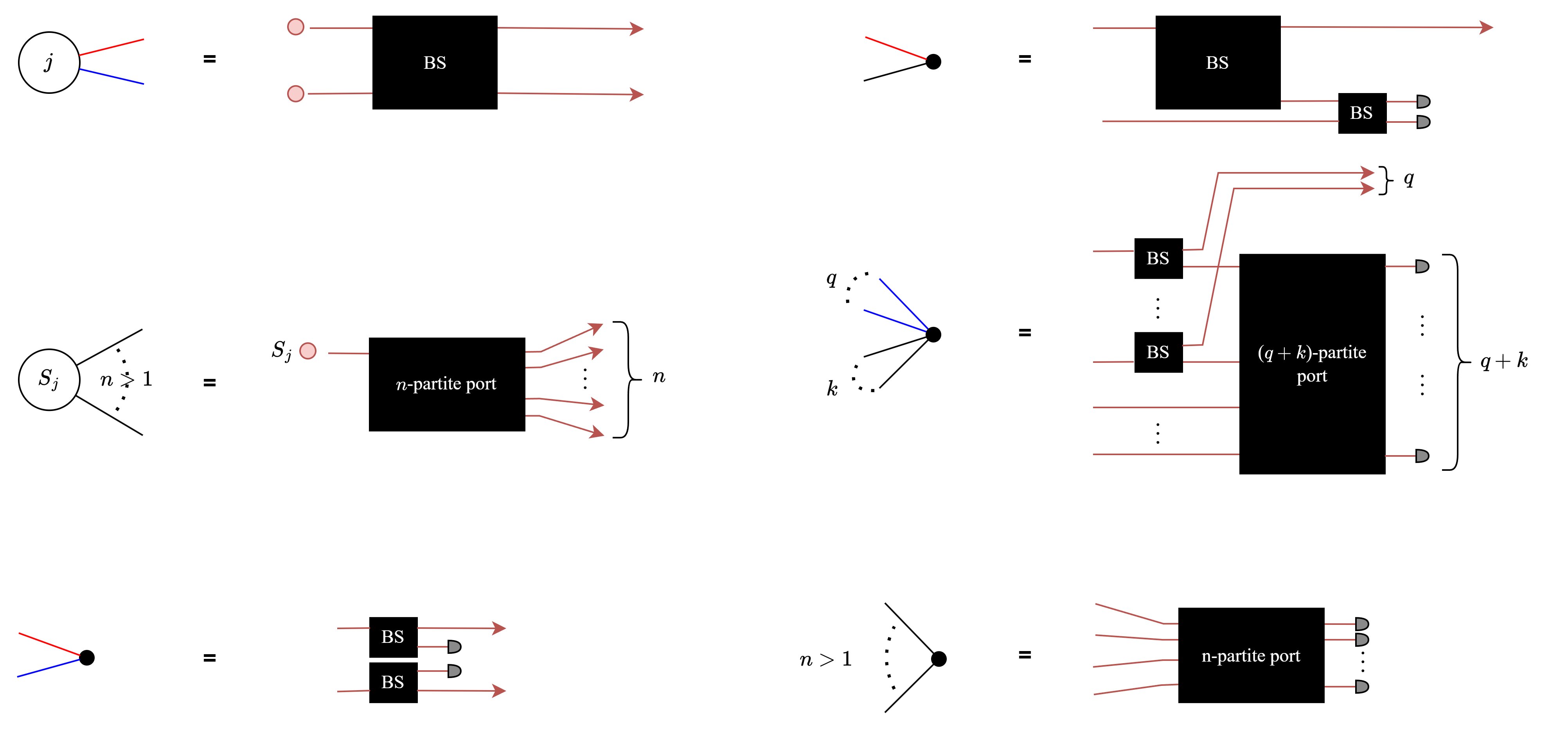}
        \caption{Translation rules from EPM bigraph elements to dual-rail linear optical elements, adapted from Fig.~2 of Ref.~\cite{chin2024heralded} with an additional rule for dots connected only by black edges. Each bigraph block (left) is mapped to its optical counterpart (right) composed of BSs and symmetric or asymmetric multiport interferometers. Assembling these blocks according to the bigraph connectivity yields the full heralded linear optical circuit.}

    \label{fig:translation_rules_dualrail}
\end{figure*}

\red{
\paragraph*{Translation rules.---}
Figure~\ref{fig:translation_rules_dualrail} summarizes the translation rules from EPM bigraph elements to dual-rail linear optical elements. Each rule maps a local circle--dot subgraph to an optical block composed of BSs and multiport interferometers. These rules apply to any EPM bigraph; to construct the circuit for a given Dicke state $\ket{D_n^k}$, one applies them to each bigraph element and assembles the resulting blocks according to the bigraph connectivity. The application of these rules to Dicke-state circuits is summarized for the general $D_n^k$ case in Fig.~\ref{fig:linear optical network blocks} 
}

\bibliographystyle{unsrt}
\bibliography{reference}

@article{bluvstein2024logical, 
year = {2024}, 
title = {{Logical quantum processor based on reconfigurable atom arrays}}, 
author = {Bluvstein, Dolev and Evered, Simon J. and Geim, Alexandra A. and Li, Sophie H. and Zhou, Hengyun and Manovitz, Tom and Ebadi, Sepehr and Cain, Madelyn and Kalinowski, Marcin and Hangleiter, Dominik and Ataides, J. Pablo Bonilla and Maskara, Nishad and Cong, Iris and Gao, Xun and Rodriguez, Pedro Sales and Karolyshyn, Thomas and Semeghini, Giulia and Gullans, Michael J. and Greiner, Markus and Vuletić, Vladan and Lukin, Mikhail D.}, 
journal = {Nature}, 
issn = {0028-0836}, 
doi = {10.1038/s41586-023-06927-3}, 
pmid = {38056497}, 
pmcid = {PMC10830422}, 
eprint = {2312.03982}, 
pages = {58--65}, 
number = {7997}, 
volume = {626}
}

@article{madsen2022quantum, 
year = {2022}, 
title = {{Quantum computational advantage with a programmable photonic processor}}, 
author = {Madsen, Lars S. and Laudenbach, Fabian and Askarani, Mohsen Falamarzi. and Rortais, Fabien and Vincent, Trevor and Bulmer, Jacob F. F. and Miatto, Filippo M. and Neuhaus, Leonhard and Helt, Lukas G. and Collins, Matthew J. and Lita, Adriana E. and Gerrits, Thomas and Nam, Sae Woo and Vaidya, Varun D. and Menotti, Matteo and Dhand, Ish and Vernon, Zachary and Quesada, Nicolás and Lavoie, Jonathan}, 
journal = {Nature}, 
issn = {0028-0836}, 
doi = {10.1038/s41586-022-04725-x}, 
pmid = {35650354}, 
pmcid = {PMC9159949}, 
pages = {75--81}, 
number = {7912}, 
volume = {606}
}

@article{proietti2021experimental, 
year = {2021}, 
title = {{Experimental quantum conference key agreement}}, 
author = {Proietti, Massimiliano and Ho, Joseph and Grasselli, Federico and Barrow, Peter and Malik, Mehul and Fedrizzi, Alessandro}, 
journal = {Science Advances}, 
doi = {10.1126/sciadv.abe0395}, 
pmid = {34088659}, 
pmcid = {PMC8177693}, 
eprint = {2002.01491}, 
pages = {eabe0395}, 
number = {23}, 
volume = {7}
}

@article{murta2020quantum, 
year = {2020}, 
title = {{Quantum Conference Key Agreement: A Review}}, 
author = {Murta, Gláucia and Grasselli, Federico and Kampermann, Hermann and Bruß, Dagmar}, 
journal = {Advanced Quantum Technologies}, 
issn = {2511-9044}, 
doi = {10.1002/qute.202000025}, 
eprint = {2003.10186}, 
number = {11}, 
volume = {3}
}

@article{guo2019advances, 
year = {2019}, 
title = {{Advances in Quantum Dense Coding}}, 
author = {Guo, Yu and Liu, Bi‐Heng and Li, Chuan‐Feng and Guo, Guang‐Can}, 
journal = {Advanced Quantum Technologies}, 
issn = {2511-9044}, 
doi = {10.1002/qute.201900011}, 
eprint = {1904.12252}, 
number = {5-6}, 
volume = {2}
}

@article{hu2018beating, 
year = {2018}, 
title = {{Beating the channel capacity limit for superdense coding with entangled ququarts}}, 
author = {Hu, Xiao-Min and Guo, Yu and Liu, Bi-Heng and Huang, Yun-Feng and Li, Chuan-Feng and Guo, Guang-Can}, 
journal = {Science Advances}, 
doi = {10.1126/sciadv.aat9304}, 
pmid = {30035231}, 
pmcid = {PMC6054506}, 
eprint = {1807.10452}, 
pages = {eaat9304}, 
number = {7}, 
volume = {4}
}

@article{al2023suppressing, 
year = {2023}, 
title = {{Suppressing quantum errors by scaling a surface code logical qubit}}, 
author = {Google Quantum AI et al.}, 
journal = {Nature}, 
issn = {0028-0836}, 
doi = {10.1038/s41586-022-05434-1}, 
pmid = {36813892}, 
pmcid = {PMC9946823}, 
eprint = {2207.06431}, 
pages = {676--681}, 
number = {7949}, 
volume = {614}
}

@article{luo2019quantum, 
year = {2019}, 
title = {{Quantum Teleportation in High Dimensions}}, 
author = {Luo, Yi-Han and Zhong, Han-Sen and Erhard, Manuel and Wang, Xi-Lin and Peng, Li-Chao and Krenn, Mario and Jiang, Xiao and Li, Li and Liu, Nai-Le and Lu, Chao-Yang and Zeilinger, Anton and Pan, Jian-Wei}, 
journal = {Physical Review Letters}, 
issn = {0031-9007}, 
doi = {10.1103/physrevlett.123.070505}, 
pmid = {31491117}, 
eprint = {1906.09697}, 
pages = {070505}, 
number = {7}, 
volume = {123}
}

@article{hermans2022qubit, 
year = {2022}, 
title = {{Qubit teleportation between non-neighbouring nodes in a quantum network}}, 
author = {Hermans, S. L. N. and Pompili, M. and Beukers, H. K. C. and Baier, S. and Borregaard, J. and Hanson, R.}, 
journal = {Nature}, 
issn = {0028-0836}, 
doi = {10.1038/s41586-022-04697-y}, 
pmid = {35614248}, 
pmcid = {PMC9132773}, 
eprint = {2110.11373}, 
pages = {663--668}, 
number = {7911}, 
volume = {605}
}

@article{horodecki2009quantum,
  title = {Quantum entanglement},
  author = {Horodecki, Ryszard and Horodecki, Pawe\l{} and Horodecki, Micha\l{} and Horodecki, Karol},
  journal = {Rev. Mod. Phys.},
  volume = {81},
  issue = {2},
  pages = {865--942},
  numpages = {0},
  year = {2009},
  month = {Jun},
  publisher = {American Physical Society},
  doi = {10.1103/RevModPhys.81.865},
  url = {https://link.aps.org/doi/10.1103/RevModPhys.81.865}
}

@article{ekert1991quantum,
  title = {Quantum cryptography based on {B}ell's theorem},
  author = {Ekert, Artur K.},
  journal = {Phys. Rev. Lett.},
  volume = {67},
  issue = {6},
  pages = {661--663},
  numpages = {0},
  year = {1991},
  month = {Aug},
  publisher = {American Physical Society},
  doi = {10.1103/PhysRevLett.67.661},
  url = {https://link.aps.org/doi/10.1103/PhysRevLett.67.661}
}

@article{Bouwmeester1999observation,
  title = {Observation of Three-Photon Greenberger-Horne-Zeilinger Entanglement},
  author = {Bouwmeester, Dik and Pan, Jian-Wei and Daniell, Matthew and Weinfurter, Harald and Zeilinger, Anton},
  journal = {Phys. Rev. Lett.},
  volume = {82},
  issue = {7},
  pages = {1345--1349},
  numpages = {0},
  year = {1999},
  month = {Feb},
  publisher = {American Physical Society},
  doi = {10.1103/PhysRevLett.82.1345},
  url = {https://link.aps.org/doi/10.1103/PhysRevLett.82.1345}
}

@Article{Lu2007experimental,
author={Lu, Chao-Yang
and Zhou, Xiao-Qi
and G{\"u}hne, Otfried
and Gao, Wei-Bo
and Zhang, Jin
and Yuan, Zhen-Sheng
and Goebel, Alexander
and Yang, Tao
and Pan, Jian-Wei},
title={Experimental entanglement of six photons in graph states},
journal={Nature Physics},
year={2007},
month={Feb},
day={01},
volume={3},
number={2},
pages={91-95},
issn={1745-2481},
doi={10.1038/nphys507},
url={https://doi.org/10.1038/nphys507}
}

@article{linington2008robust,
  title = {Robust creation of arbitrary-sized {D}icke states of trapped ions by global addressing},
  author = {Linington, I. E. and Vitanov, N. V.},
  journal = {Phys. Rev. A},
  volume = {77},
  issue = {1},
  pages = {010302},
  numpages = {4},
  year = {2008},
  month = {Jan},
  publisher = {American Physical Society},
  doi = {10.1103/PhysRevA.77.010302},
  url = {https://link.aps.org/doi/10.1103/PhysRevA.77.010302}
}

@article{hume2009preparation,
  title = {Preparation of {D}icke states in an ion chain},
  author = {Hume, D. B. and Chou, C. W. and Rosenband, T. and Wineland, D. J.},
  journal = {Phys. Rev. A},
  volume = {80},
  issue = {5},
  pages = {052302},
  numpages = {5},
  year = {2009},
  month = {Nov},
  publisher = {American Physical Society},
  doi = {10.1103/PhysRevA.80.052302},
  url = {https://link.aps.org/doi/10.1103/PhysRevA.80.052302}
}

@article{ivanov2013creation,
doi = {10.1088/1367-2630/15/2/023039},
url = {https://dx.doi.org/10.1088/1367-2630/15/2/023039},
year = {2013},
month = {feb},
publisher = {IOP Publishing},
volume = {15},
number = {2},
pages = {023039},
author = {Ivanov, Svetoslav S and Vitanov, Nikolay V and Korolkova, Natalia V},
title = {Creation of arbitrary {D}icke and {NOON} states of trapped-ion qubits by global addressing with composite pulses},
journal = {New Journal of Physics},
}

@article{stockton2004deterministic,
  title = {Deterministic Dicke-state preparation with continuous measurement and control},
  author = {Stockton, John K. and van Handel, Ramon and Mabuchi, Hideo},
  journal = {Phys. Rev. A},
  volume = {70},
  issue = {2},
  pages = {022106},
  numpages = {11},
  year = {2004},
  month = {Aug},
  publisher = {American Physical Society},
  doi = {10.1103/PhysRevA.70.022106},
  url = {https://link.aps.org/doi/10.1103/PhysRevA.70.022106}
}

@article{xiao2007generation,
  title = {Generation of atomic entangled states with selective resonant interaction in cavity quantum electrodynamics},
  author = {Xiao, Yun-Feng and Zou, Xu-Bo and Guo, Guang-Can},
  journal = {Phys. Rev. A},
  volume = {75},
  issue = {1},
  pages = {012310},
  numpages = {5},
  year = {2007},
  month = {Jan},
  publisher = {American Physical Society},
  doi = {10.1103/PhysRevA.75.012310},
  url = {https://link.aps.org/doi/10.1103/PhysRevA.75.012310}
}

@article{bennett1992communication,
  title = {Communication via one- and two-particle operators on {E}instein-{P}odolsky-{R}osen states},
  author = {Bennett, Charles H. and Wiesner, Stephen J.},
  journal = {Phys. Rev. Lett.},
  volume = {69},
  issue = {20},
  pages = {2881--2884},
  numpages = {0},
  year = {1992},
  month = {Nov},
  publisher = {American Physical Society},
  doi = {10.1103/PhysRevLett.69.2881},
  url = {https://link.aps.org/doi/10.1103/PhysRevLett.69.2881}
}

@article{Ozdemir2007necessary,
doi = {10.1088/1367-2630/9/2/043},
url = {https://dx.doi.org/10.1088/1367-2630/9/2/043},
year = {2007},
month = {feb},
publisher = {},
volume = {9},
number = {2},
pages = {43},
author = {Özdemir, S K and Shimamura, J and Imoto, N},
title = {A necessary and sufficient condition to play games in quantum mechanical settings},
journal = {New Journal of Physics},
}

@article{bennett1993teleporting,
  title = {Teleporting an unknown quantum state via dual classical and Einstein-Podolsky-Rosen channels},
  author = {Bennett, Charles H. and Brassard, Gilles and Cr\'epeau, Claude and Jozsa, Richard and Peres, Asher and Wootters, William K.},
  journal = {Phys. Rev. Lett.},
  volume = {70},
  issue = {13},
  pages = {1895--1899},
  numpages = {0},
  year = {1993},
  month = {Mar},
  publisher = {American Physical Society},
  doi = {10.1103/PhysRevLett.70.1895},
  url = {https://link.aps.org/doi/10.1103/PhysRevLett.70.1895}
}

@article{Shao2010deterministic,
doi = {10.1209/0295-5075/90/50003},
url = {https://dx.doi.org/10.1209/0295-5075/90/50003},
year = {2010},
month = {jun},
publisher = {},
volume = {90},
number = {5},
pages = {50003},
author = {Shao, Xiao-Qiang and Chen, Li and Zhang, Shou and Zhao, Yong-Fang and Yeon, Kyu-Hwang},
title = {Deterministic generation of arbitrary multi-atom symmetric {D}icke states by a combination of quantum Zeno dynamics and adiabatic passage},
journal = {Europhysics Letters}
}

@article{wu2017generation,
  title = {Generation of Dicke states in the ultrastrong-coupling regime of circuit QED systems},
  author = {Wu, Chunfeng and Guo, Chu and Wang, Yimin and Wang, Gangcheng and Feng, Xun-Li and Chen, Jing-Ling},
  journal = {Phys. Rev. A},
  volume = {95},
  issue = {1},
  pages = {013845},
  numpages = {7},
  year = {2017},
  month = {Jan},
  publisher = {American Physical Society},
  doi = {10.1103/PhysRevA.95.013845},
  url = {https://link.aps.org/doi/10.1103/PhysRevA.95.013845}
}

@article{tutteshort1954, 
year = {1954}, 
title = {{A short proof of the factor theorem for finite graphs}}, 
author = {Tutte, W. T.}, 
journal = {Canadian Journal of Mathematics}, 
issn = {0008-414X}, 
doi = {10.4153/cjm-1954-033-3}, 
pages = {347--352}, 
number = {0}, 
volume = {6}
}

@article{chin2024heralded, 
year = {2024}, 
title = {{Heralded Optical Entanglement Generation via the Graph Picture of Linear Quantum Networks}}, 
author = {Chin, Seungbeom and Karczewski, Marcin and Kim, Yong-Su}, 
journal = {Quantum}, 
doi = {10.22331/q-2024-12-18-1572}, 
eprint = {2310.10291},  
pages = {1572}, 
volume = {8}, 
}

@article{chin2024shortcut, 
year = {2024}, 
title = {{Shortcut to multipartite entanglement generation: A graph approach to boson subtractions}}, 
author = {Chin, Seungbeom and Kim, Yong-Su and Karczewski, Marcin}, 
journal = {npj Quantum Information}, 
doi = {10.1038/s41534-024-00845-6}, 
eprint = {2211.04042}, 
pages = {67}, 
number = {1}, 
volume = {10}, 
}

@article{chin2021graph, 
year = {2021}, 
title = {{Graph Picture of Linear Quantum Networks and Entanglement}}, 
author = {Chin, Seungbeom and Kim, Yong-Su and Lee, Sangmin}, 
journal = {Quantum}, 
doi = {10.22331/q-2021-12-23-611}, 
eprint = {2101.00392}, 
pages = {611}, 
volume = {5}
}

@article{chin2024boson, 
year = {2023}, 
title = {{Efficient Photonic Graph State Generation}}, 
author = {Chin, Seungbeom and Munro, William J.}, 
journal={arXiv preprint arXiv:2306.15148},
year={2023}
}

@article{dicke1954coherence,
  title = {Coherence in Spontaneous Radiation Processes},
  author = {Dicke, R. H.},
  journal = {Phys. Rev.},
  volume = {93},
  issue = {1},
  pages = {99--110},
  numpages = {0},
  year = {1954},
  month = {Jan},
  publisher = {American Physical Society},
  doi = {10.1103/PhysRev.93.99},
  url = {https://link.aps.org/doi/10.1103/PhysRev.93.99}
}

@article{prevedel2009experimental,
  title = {Experimental Realization of {D}icke States of up to Six Qubits for Multiparty Quantum Networking},
  author = {Prevedel, R. and Cronenberg, G. and Tame, M. S. and Paternostro, M. and Walther, P. and Kim, M. S. and Zeilinger, A.},
  journal = {Phys. Rev. Lett.},
  volume = {103},
  issue = {2},
  pages = {020503},
  numpages = {4},
  year = {2009},
  month = {Jul},
  publisher = {American Physical Society},
  doi = {10.1103/PhysRevLett.103.020503},
  url = {https://link.aps.org/doi/10.1103/PhysRevLett.103.020503}
}

@article{wieczorek2009experimental,
  title = {Experimental Entanglement of a Six-Photon Symmetric Dicke State},
  author = {Wieczorek, Witlef and Krischek, Roland and Kiesel, Nikolai and Michelberger, Patrick and T\'oth, G\'eza and Weinfurter, Harald},
  journal = {Phys. Rev. Lett.},
  volume = {103},
  issue = {2},
  pages = {020504},
  numpages = {4},
  year = {2009},
  month = {Jul},
  publisher = {American Physical Society},
  doi = {10.1103/PhysRevLett.103.020504},
  url = {https://link.aps.org/doi/10.1103/PhysRevLett.103.020504}
}

@article{ouyang2021robust,
  author={Ouyang, Yingkai and Shettell, Nathan and Markham, Damian},
  journal={IEEE Transactions on Information Theory}, 
  title={Robust Quantum Metrology With Explicit Symmetric States}, 
  year={2022},
  volume={68},
  number={3},
  pages={1809-1821},
  doi={10.1109/TIT.2021.3132634}}

@article{saleem2024achieving,
  title = {Achieving the {H}eisenberg limit with {D}icke states in noisy quantum metrology},
  author = {Saleem, Zain H. and Perlin, Michael and Shaji, Anil and Gray, Stephen K.},
  journal = {Phys. Rev. A},
  volume = {109},
  issue = {5},
  pages = {052615},
  numpages = {10},
  year = {2024},
  month = {May},
  publisher = {American Physical Society},
  doi = {10.1103/PhysRevA.109.052615},
  url = {https://link.aps.org/doi/10.1103/PhysRevA.109.052615}
}

@article{zhao2011efficient,
title = {Efficient scheme for the preparation of symmetric {D}icke states via cross-Kerr nonlinearity},
journal = {Physics Letters A},
volume = {375},
number = {3},
pages = {401-405},
year = {2011},
issn = {0375-9601},
doi = {https://doi.org/10.1016/j.physleta.2010.11.028},
url = {https://www.sciencedirect.com/science/article/pii/S0375960110014970},
author = {Chunran Zhao and Liu Ye},
}

@article{toth2012multipartite,
  title = {Multipartite entanglement and high-precision metrology},
  author = {T\'oth, G\'eza},
  journal = {Phys. Rev. A},
  volume = {85},
  issue = {2},
  pages = {022322},
  numpages = {8},
  year = {2012},
  month = {Feb},
  publisher = {American Physical Society},
  doi = {10.1103/PhysRevA.85.022322},
  url = {https://link.aps.org/doi/10.1103/PhysRevA.85.022322}
}

@article{lamata2013deterministic,
  title = {Deterministic generation of arbitrary symmetric states and entanglement classes},
  author = {Lamata, L. and L\'opez, C. E. and Lanyon, B. P. and Bastin, T. and Retamal, J. C. and Solano, E.},
  journal = {Phys. Rev. A},
  volume = {87},
  issue = {3},
  pages = {032325},
  numpages = {5},
  year = {2013},
  month = {Mar},
  publisher = {American Physical Society},
  doi = {10.1103/PhysRevA.87.032325},
  url = {https://link.aps.org/doi/10.1103/PhysRevA.87.032325}
}

@article{kiesel2007experimental,
  title = {Experimental Observation of Four-Photon Entangled {D}icke State with High Fidelity},
  author = {Kiesel, N. and Schmid, C. and T\'oth, G. and Solano, E. and Weinfurter, H.},
  journal = {Phys. Rev. Lett.},
  volume = {98},
  issue = {6},
  pages = {063604},
  numpages = {4},
  year = {2007},
  month = {Feb},
  publisher = {American Physical Society},
  doi = {10.1103/PhysRevLett.98.063604},
  url = {https://link.aps.org/doi/10.1103/PhysRevLett.98.063604}
}

@article{leibfried2005creation,
  title={Creation of a six-atom ‘{S}chr{\"o}dinger cat’state},
  author={Leibfried, Dietrich and Knill, Emanuel and Seidelin, Signe and Britton, Joe and Blakestad, R Brad and Chiaverini, John and Hume, David B and Itano, Wayne M and Jost, John D and Langer, Christopher and others},
  journal={Nature},
  volume={438},
  number={7068},
  pages={639--642},
  year={2005},
  publisher={Nature Publishing Group UK London}
}

@article{hume2007high,
  title={High-Fidelity Adaptive Qubit Detection through Repetitive Quantum Nondemolition Measurements},
  author={Hume, DB and Rosenband, Till and Wineland, David J},
  journal={Physical review letters},
  volume={99},
  number={12},
  pages={120502},
  year={2007},
  publisher={APS}
}

@article{monz201114,
  title={14-qubit entanglement: Creation and coherence},
  author={Monz, Thomas and Schindler, Philipp and Barreiro, Julio T and Chwalla, Michael and Nigg, Daniel and Coish, William A and Harlander, Maximilian and H{\"a}nsel, Wolfgang and Hennrich, Markus and Blatt, Rainer},
  journal={Physical Review Letters},
  volume={106},
  number={13},
  pages={130506},
  year={2011},
  publisher={APS}
}

@article{mandel2003controlled,
  title={Controlled collisions for multi-particle entanglement of optically trapped atoms},
  author={Mandel, Olaf and Greiner, Markus and Widera, Artur and Rom, Tim and H{\"a}nsch, Theodor W and Bloch, Immanuel},
  journal={Nature},
  volume={425},
  number={6961},
  pages={937--940},
  year={2003},
  publisher={Nature Publishing Group UK London}
}

@article{bernien2017probing,
  title={Probing many-body dynamics on a 51-atom quantum simulator},
  author={Bernien, Hannes and Schwartz, Sylvain and Keesling, Alexander and Levine, Harry and Omran, Ahmed and Pichler, Hannes and Choi, Soonwon and Zibrov, Alexander S and Endres, Manuel and Greiner, Markus and others},
  journal={Nature},
  volume={551},
  number={7682},
  pages={579--584},
  year={2017},
  publisher={Nature Publishing Group UK London}
}

@article{omran2019generation,
  title={Generation and manipulation of {S}chr{\"o}dinger cat states in {R}ydberg atom arrays},
  author={Omran, Ahmed and Levine, Harry and Keesling, Alexander and Semeghini, Giulia and Wang, Tout T and Ebadi, Sepehr and Bernien, Hannes and Zibrov, Alexander S and Pichler, Hannes and Choi, Soonwon and others},
  journal={Science},
  volume={365},
  number={6453},
  pages={570--574},
  year={2019},
  publisher={American Association for the Advancement of Science}
}

@article{neeley2010generation,
  title={Generation of three-qubit entangled states using superconducting phase qubits},
  author={Neeley, Matthew and Bialczak, Radoslaw C and Lenander, M and Lucero, Erik and Mariantoni, Matteo and O’connell, AD and Sank, D and Wang, H and Weides, M and Wenner, J and others},
  journal={Nature},
  volume={467},
  number={7315},
  pages={570--573},
  year={2010},
  publisher={Nature Publishing Group UK London}
}

@article{dicarlo2010preparation,
  title={Preparation and measurement of three-qubit entanglement in a superconducting circuit},
  author={DiCarlo, Leonardo and Reed, Matthew D and Sun, Luyan and Johnson, Blake R and Chow, Jerry M and Gambetta, Jay M and Frunzio, Luigi and Girvin, Steven M and Devoret, Michel H and Schoelkopf, Robert J},
  journal={Nature},
  volume={467},
  number={7315},
  pages={574--578},
  year={2010},
  publisher={Nature Publishing Group UK London}
}

@article{song201710,
  title={10-qubit entanglement and parallel logic operations with a superconducting circuit},
  author={Song, Chao and Xu, Kai and Liu, Wuxin and Yang, Chui-ping and Zheng, Shi-Biao and Deng, Hui and Xie, Qiwei and Huang, Keqiang and Guo, Qiujiang and Zhang, Libo and others},
  journal={Physical review letters},
  volume={119},
  number={18},
  pages={180511},
  year={2017},
  publisher={APS}
}

@article{lee2022entangling,
  title={Entangling three identical particles via spatial overlap},
  author={Lee, Donghwa and Pramanik, Tanumoy and Hong, Seongjin and Cho, Young-Wook and Lim, Hyang-Tag and Chin, Seungbeom and Kim, Yong-Su},
  journal={Optics Express},
  volume={30},
  number={17},
  pages={30525--30535},
  year={2022},
  publisher={Optica Publishing Group}
}

@article{dur2000three,
  title={Three qubits can be entangled in two inequivalent ways},
  author={D{\"u}r, Wolfgang and Vidal, Guifre and Cirac, J Ignacio},
  journal={Physical Review A},
  volume={62},
  number={6},
  pages={062314},
  year={2000},
  publisher={APS}
}

@article{ma2024multipartite,
  title={Multipartite entanglement measures: A review},
  author={Ma, Mengru and Li, Yinfei and Shang, Jiangwei},
  journal={Fundamental Research},
  year={2024},
  publisher={Elsevier}
}

@article{scursulim2025multiclass,
  title={Multiclass {P}ortfolio {O}ptimization via {V}ariational {Q}uantum {E}igensolver with {D}icke {S}tate {A}nsatz},
  author={Scursulim, Jos{\'e} Victor S and Langeloh, Gabriel Mattos and Beltran, Victor Leme and Brito, Samura{\'\i}},
  journal={arXiv preprint arXiv:2508.13954},
  year={2025}
}

@article{Kasture2018,
	author = {Kasture, Sachin},
	title = {{Scalable approach to generation of large symmetric {D}icke states}},
	journal = {Phys. Rev. A},
	volume = {97},
	number = {4},
	pages = {043862},
	year = {2018},
	month = {Apr},
	issn = {2469-9934},
	publisher = {American Physical Society},
	doi = {10.1103/PhysRevA.97.043862}
}

@article{chin2024exponentially,
  title={Exponentially enhanced scheme for the heralded qudit {G}reenberger-{H}orne-{Z}eilinger state in linear optics},
  author={Chin, Seungbeom and Ryu, Junghee and Kim, Yong-Su},
  journal={Physical Review Letters},
  volume={133},
  number={25},
  pages={253601},
  year={2024},
  publisher={APS}
}

@article{chin2024creating,
  title={Creating highly symmetric qudit heralded entanglement through highly symmetric graphs},
  author={Chin, Seungbeom},
  journal={arXiv preprint arXiv:2404.05273},
  year={2024}
}

@article{bustard2024toward,
  title={Toward deterministic sources: Photon generation in a fiber-cavity quantum memory},
  author={Bustard, Philip J and Tannous, Ramy and Bonsma-Fisher, Kent and Poitras, Daniel and Hnatovsky, Cyril and Mihailov, Stephen J and England, Duncan and Sussman, Benjamin J},
  journal={Physical Review A},
  volume={109},
  number={1},
  pages={013711},
  year={2024},
  publisher={APS}
}

@article{meyer2020single,
  title={Single-photon sources: Approaching the ideal through multiplexing},
  author={Meyer-Scott, Evan and Silberhorn, Christine and Migdall, Alan},
  journal={Review of Scientific Instruments},
  volume={91},
  number={4},
  year={2020},
  publisher={AIP Publishing}
}

@article{walter2016multipartite,
  title={Multipartite entanglement},
  author={Walter, Michael and Gross, David and Eisert, Jens},
  journal={Quantum Information: From Foundations to Quantum Technology Applications},
  pages={293--330},
  year={2016},
  publisher={Wiley Online Library}
}

@article{Ouyang14Permutation,
  title = {Permutation-invariant quantum codes},
  author = {Ouyang, Yingkai},
  journal = {Phys. Rev. A},
  volume = {90},
  issue = {6},
  pages = {062317},
  numpages = {11},
  year = {2014},
  month = {Dec},
  publisher = {American Physical Society},
  doi = {10.1103/PhysRevA.90.062317},
  url = {https://link.aps.org/doi/10.1103/PhysRevA.90.062317}
}

@article{Aydin2024familyof,
  doi = {10.22331/q-2024-04-30-1321},
  url = {https://doi.org/10.22331/q-2024-04-30-1321},
  title = {A family of permutationally invariant quantum codes},
  author = {Aydin, Arda and Alekseyev, Max A. and Barg, Alexander},
  journal = {{Quantum}},
  issn = {2521-327X},
  publisher = {{Verein zur F{\"{o}}rderung des Open Access Publizierens in den Quantenwissenschaften}},
  volume = {8},
  pages = {1321},
  month = apr,
  year = {2024}
}

@article{forbes2025heralded,
  title={Heralded generation of entanglement with photons},
  author={Forbes, Imogen and Ghafari, Farzad and Deacon, Edward CR and Singh, Sukhjit P and Lavie, Emilien and Yard, Patrick and Shaw, Reece D and Laing, Anthony and Tischler, Nora},
  journal={Reports on Progress in Physics},
  volume={88},
  number={8},
  pages={086002},
  year={2025},
  publisher={IOP Publishing}
}

\end{document}